\documentclass[lettersize,journal]{IEEEtran}
 \usepackage{cite}
\usepackage{xcolor}
\usepackage{tcolorbox}
\tcbuselibrary{listings}

\newtcblisting{Python}[1][]{
 colback=white,
 colframe=black,
 listing only,
 listing options={
 language=Python,
 basicstyle=\footnotesize\ttfamily, 
 keywordstyle=\color{blue},
 commentstyle=\color{green!40!black},
 stringstyle=\color{orange},
 breaklines=true,
 showstringspaces=false,
 tabsize=2,
 numbers=left,
 numberstyle=\tiny\color{gray},
 #1 
 },
 left=6mm,
 top=0mm, 
 bottom=0mm, 
 boxsep=0mm, 
 arc=0mm, 
 boxrule=0.2mm, 
 overlay={
 \begin{tcbclipinterior}
 \fill[gray!25] (frame.south west) rectangle ([xshift=6mm]frame.north west);
 \end{tcbclipinterior}
 }
}

\usepackage{enumitem}
\usepackage{amsfonts}
\usepackage{enumitem,kantlipsum}
\usepackage{textcomp}
\usepackage{stfloats}
\usepackage{url}
\usepackage{caption}
\usepackage{graphics}
\usepackage{verbatim}
\usepackage{pifont}
\usepackage{float}
\usepackage{amsmath}
\usepackage{cancel}
\usepackage{amsmath}
\usepackage{graphicx}
\usepackage{wrapfig}
\usepackage{float}
\usepackage{lscape}
\usepackage{ltxtable}
\usepackage{amsmath}
\usepackage{algorithm}
\usepackage{algpseudocode}
\usepackage{amsfonts}
\usepackage{amsfonts}
\usepackage{tikz}
\usetikzlibrary{positioning, arrows.meta, shapes.geometric}
\usetikzlibrary{positioning, shapes, arrows, fit}
\usepackage{hhline}
\usepackage{caption}
\usepackage{wasysym}
\usepackage{xurl}
\usepackage{rotating}
\usepackage{graphicx}
\usepackage{bm}
\usepackage{caption}
\usepackage{array}
\usepackage{listings}
\usepackage{subcaption}
\usepackage{color} 
\usepackage{url}

\usepackage[
    colorlinks=false,
    pdfborder={0 0 0}
]{hyperref}
\usepackage{array} 
\usepackage[export]{adjustbox}
\usepackage{colortbl}
\usepackage{verbatim}
\usepackage{tikz}
\usetikzlibrary{shapes,arrows}
\usetikzlibrary{intersections}
\usetikzlibrary{automata, arrows.meta, positioning}
\usetikzlibrary{decorations.pathreplacing}
\usepackage{amsthm}
\usepackage{booktabs}
\usepackage{amsmath}
\usepackage{inputenc}
\usepackage{adjustbox}
\usepackage{url} 
\usepackage{pgfplots}
\usepackage{pgfplotstable}
\usepackage{booktabs}
\usepackage{array}
\pgfplotsset{compat=1.18}

\newcommand{\legendcolorbox}[1]{\tikz[baseline=-0.5ex]\draw[fill=#1, draw=black] (0,-0.4em) rectangle (0.9em,0.5em);}

\usepackage{multirow}
\usepackage{color,soul}
\usepackage{lipsum}
\usepackage{mathtools}
\usepackage{subcaption}
\usepackage{cuted}
\usepackage{wasysym}
\usepackage{pgfplotstable}
\usepackage{mdframed}
 \usepackage{mhchem}

\newtheorem{thm}{Theorem}

\theoremstyle{definition}

\newtheorem{game}{Game}
\newtheoremstyle{remarkbold}   
  {\topsep}                    
  {\topsep}                    
  {\normalfont}                
  {}                           
  {\bfseries}                  
  {.}                          
  { }                          
  {}                           

\theoremstyle{remarkbold}
\newtheorem*{remark}{Remark}

\usepackage[normalem]{ulem}
\usepackage[acronym]{glossaries}
\newlist{MyItemize}{itemize}{1}
\setlist[MyItemize]{
 label={\textbullet},
 leftmargin=*,
 topsep=1ex,
 partopsep=0ex,
 parsep=0ex,
 itemsep=0ex,
}

\usetikzlibrary{arrows.meta}
\usetikzlibrary{ calc}
\usepackage{smartdiagram}
\definecolor{lemon}{rgb}{1.0, 1.0, 0.13}
\usetikzlibrary{arrows.meta}
\usepackage{smartdiagram}
\usetikzlibrary{positioning, fit, calc, shapes, arrows,trees}

\usepackage[cmintegrals]{newtxmath}

\usepackage{hhline}
\definecolor{columbiablue}{rgb}{0.61, 0.87, 1.0}

\tikzset{%
 thick arrow/.style={
 -{Triangle[angle=120:1pt 1]},
 line width=0.8cm, 
 draw=blue!20
 },
 arrow label/.style={
 text=black,
 align=center
 },
 set mark/.style={
 insert path={
 node [midway, arrow label, node contents=#1]
 }
 }
}
\newcommand\deleted{\bgroup\markoverwith{\textcolor{red}{\rule[0.5ex]{2pt}{0.4pt}}}\ULon}

\usetikzlibrary{pgfplots.groupplots}

\newcommand\doublecheck{\textcolor{black}{\checkmark\kern-0em\checkmark}}
\newcommand\semidoublecheck{\textcolor{black}{\checkmark\kern-0em\bcancel{\checkmark}}}

\definecolor{lemon}{rgb}{1.0, 1.0, 0.13}

\newacronym{2FA}{2FA}{Two-Factor Authentication}
\newacronym{DBA}{DBA}{Dynamic Time Warping Barycenter Averaging}
\newacronym{DTW}{DTW}{Dynamic Time Warping}
\newacronym{FL}{FL}{Federated Learning}
\newacronym{PFL}{PFL}{Personalized Federated Learning}
\newacronym{FLF}{FLF}{Federated Learning Framework}
\newacronym{FLRBA2}{FL-RBA\textsuperscript{2}}{Federated Learning framework for Risk-Based Adaptive Authentication}
\newacronym{FPR}{FPR}{False Positive Rate}
\newacronym{GPS}{GPS}{Global Positioning System}
\newacronym{IID}{IID}{Independent and Identically Distributed}
\newacronym{LTH}{LTH}{Lower Threshold}
\newacronym{MAC}{MAC}{Message Authentication Code}
\newacronym{MAD}{MAD}{Median Absolute Deviation}
\newacronym{MCD}{MCD}{Minimum Covariance Determinant}
\newacronym{MFA}{MFA}{Multi-Factor Authentication}
\newacronym{ML}{ML}{Machine Learning}
\newacronym{MSE}{MSE}{Mean Squared Error}
\newacronym{NonIID}{Non-IID}{Non-Independent and Identically Distributed}
\newacronym{RBA}{RBA}{Risk-Based Adaptive Authentication}
\newacronym{RL}{RL}{Reinforcement Learning}
\newacronym{ROM}{ROM}{Random Oracle Model}
\newacronym{RTT}{RTT}{Round-Trip Time}
\newacronym{SFA}{SFA}{Single-Factor Authentication}
\newacronym{TPR}{TPR}{True Positive Rate}
\newacronym{UTH}{UTH}{Upper Threshold}
\newacronym{ZKP}{ZKP}{Zero-Knowledge Proof}
\newacronym{nonIID}{non-IID}{Non-Independent and Identically Distributed}
\newacronym{DP}{DP}{Differential Privacy}
\newacronym{PPT}{PPT}{Probabilistic Polynomial Time}
\newacronym{MPC}{MPC}{Multi-Party Computation}
\begin{document}

\title{Privacy-Preserving Federated Learning Framework for Risk-Based Adaptive Authentication}

\author{Yaser Baseri,
 Abdelhakim Senhaji Hafid,  
Dimitrios Makrakis and Hamidreza Fereidouni
 
\IEEEcompsocitemizethanks{\IEEEcompsocthanksitem Yaser Baseri,   Abdelhakim Senhaji Hafid, and Hamidreza Fereidouni are with  the Department of Computer Science and Operations Research, University of Montreal, Canada.
Emails: yaser.baseri@umontreal.ca; ahafid@iro.umontreal.ca; hamidreza.fereidouni@umontreal.ca.  Dimitrios Makrakis is with the School of Electrical Engineering and Computer Science, University of Ottawa, Canada. E-mail: dmakraki@uottawa.ca.\protect} 
\thanks{This work was supported by \href{https://www.nserc-crsng.gc.ca/}{NSERC} and \href{https://flexgroups.com}{Flex Group}.}}

\markboth{}%
{Baseri \MakeLowercase{\textit{et al.}}: Privacy-Preserving Federated Learning Framework for Risk-Based Adaptive Authentication}

\maketitle

\begin{abstract}
Balancing robust security with strong privacy guarantees is critical for  \gls{RBA}, particularly in decentralized settings. \gls{FL} offers a promising solution by enabling collaborative risk assessment without centralizing user data. However, existing FL approaches struggle with \gls{NonIID} user features, resulting in biased, unstable, and poorly generalized global models. This paper introduces \acrshort{FLRBA2}, a novel \acrlong{FLRBA2} 
that addresses \gls{NonIID} challenges through a mathematically grounded similarity transformation. By converting heterogeneous user features---including behavioral, biometric, contextual, interaction-based, and knowledge-based modalities---into IID similarity vectors, \acrshort{FLRBA2} supports unbiased aggregation and personalized risk modeling across distributed clients.
The framework mitigates cold-start limitations via clustering-based risk labeling, incorporates \gls{DP} to safeguard sensitive information, and employs \glspl{MAC} to ensure model integrity and authenticity. Federated updates are securely aggregated into a global model, achieving strong balance between user privacy, scalability, and adaptive authentication robustness. Rigorous game-based security proofs in the Random Oracle Model formally establish privacy, correctness, and adaptive security guarantees. Extensive experiments on keystroke, mouse, and contextual datasets validate \acrshort{FLRBA2}'s effectiveness in high-risk user detection and its resilience to model inversion and inference attacks, even under strong \gls{DP} constraints.

\end{abstract}

\IEEEpeerreviewmaketitle

\section{Introduction}

\IEEEPARstart{A}{uthentication} is a crucial component for secure access to systems and data, ensuring only authorized individuals gain entry. Traditional authentication methods, such as username/password combinations, static passcodes, and security questions, often rely on static mechanisms that apply a uniform security level regardless of context or varying risk~\cite{10947334,9432950,10654322,10916520}. This approach can lead to suboptimal security and user experience.
In contrast, adaptive authentication dynamically adjusts authentication strength based on real-time risk assessments. By continuously evaluating factors like user location, device type, and recent login history, it intelligently selects appropriate authentication factors, such as \gls{2FA}~\cite{10818582} or \gls{MFA}~\cite{9432950,10916520}. This method proactively enhances security by applying stronger measures when necessary, while simultaneously improving user experience by minimizing friction for legitimate users in low-risk situations.

\gls{RBA} has emerged as a prominent approach to balance security and user convenience. It dynamically assesses the risk of each login attempt by considering  the user features (knowledge-based, biometric, behavioral, contextual, and interaction-based), tailoring verification requirements in real-time based on the calculated risk level~\cite{wiefling2020more}. This enhances user experience by only requiring additional steps for suspicious activity, such as a login from an unrecognized device~\cite{sepczuk2018new}. \gls{RBA} categorizes risks into levels (e.g., low, medium, high), prompting users for credentials according to the assigned risk. For instance, low-risk scenarios might only require a password, while higher risk levels necessitate more sophisticated authentication methods.

\begin{figure}
 \centering
 \Large
 \resizebox{\linewidth}{!}{
 \begin{tikzpicture}[scale=1,
 node distance=3cm and 4cm,
 every node/.style={align=center},
 arrow/.style={thick,->,>=Stealth}
 ]

 \node (user) [draw=none, fill=none] {
 \includegraphics[width=1.5cm]{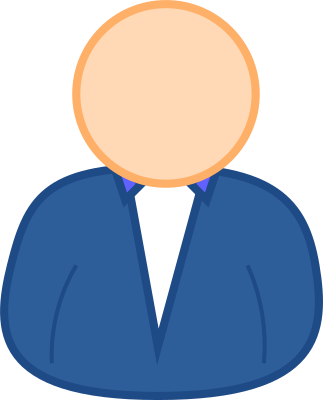}
 };
 \node [below=0.2cm of user] {User};

 \node (request) [right=of user, draw, minimum width=3cm, minimum height=2cm, align=center,fill=blue!10,rectangle, draw, rounded corners, text centered] {Initial \\Access\\ Request};

 \node (data) [right=of request, draw, minimum width=3cm, minimum height=2cm, align=center,fill=blue!10,rectangle, draw, rounded corners, text centered] {Comprehensive\\Data\\Collection};

 \node (device) [right=of data, draw, minimum width=3cm, minimum height=2cm, align=center,fill=blue!10,rectangle, draw, rounded corners, text centered] {Device and\\Media\\Recognition};

 \node (policies) [below=of request, yshift=1.5cm, draw, minimum width=3cm, minimum height=2cm, align=center,fill=blue!10,rectangle, draw, rounded corners, text centered] {Policies\\and\\Rules};

 \node (LowRisk) [below=0.8cm of policies, draw, minimum width=3cm, minimum height=1cm, align=center,fill=green!10,rectangle, draw, rounded corners, text centered] {Low Risk};
 \node (MediumRisk) [below=0.8cm of LowRisk, draw, minimum width=3cm, minimum height=1cm, align=center,fill=yellow!10,rectangle, draw, rounded corners, text centered, yshift=-0.4cm] {Medium Risk};
 \node (HighRisk) [below=0.8cm of MediumRisk, draw, minimum width=3cm, minimum height=1cm, align=center,fill=red!10,rectangle, draw, rounded corners, text centered, yshift=-0.4cm] {High Risk};

 \node (assessment) [left=of MediumRisk, xshift=2.5cm, draw, minimum width=3cm, minimum height=2cm, align=center,fill=blue!10,rectangle, draw, rounded corners, text centered] {Intelligent\\Risk\\Assessment};

 \node (SimpleAuth) [right=of LowRisk, draw, align=center, minimum width=3.5cm, xshift=-2.7cm] {
 \includegraphics[height=1cm,width=1cm]{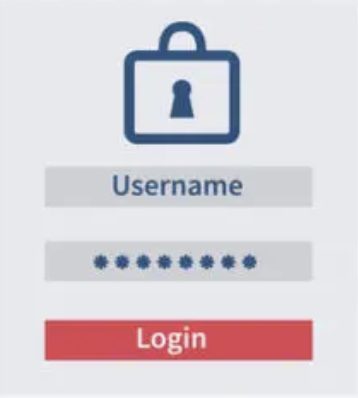} \\ Simple Auth.
 };
 \node (MoreSteps) [right=of MediumRisk, align=center, minimum width=3.5cm, xshift=-2.7cm, draw] {
 \includegraphics[height=1cm,width=1cm]{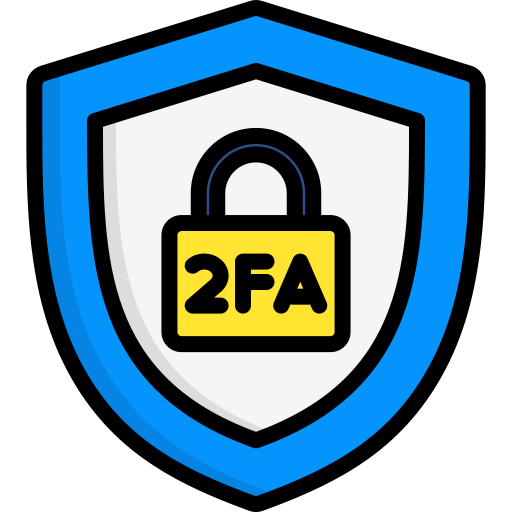} \\ More Steps
 };
 \node (AdvancedAuth) [right=of HighRisk, align=center, minimum width=3.5cm, xshift=-2.7cm, draw] {
 \includegraphics[height=1cm,width=1cm]{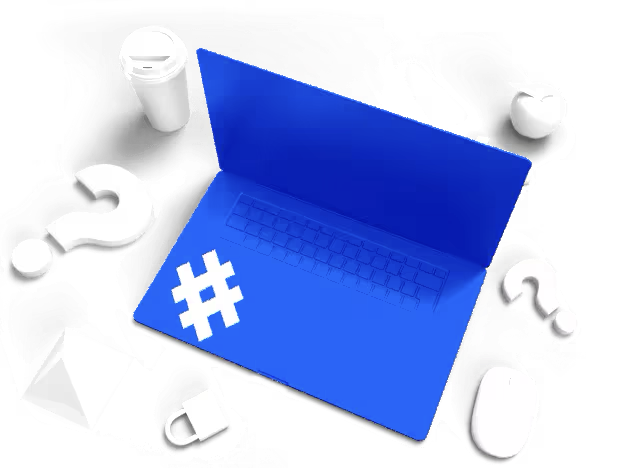} \\ Advanced Auth.
 };

 \node (UserVerification) [below=1cm of AdvancedAuth, xshift=2.7cm, draw, minimum width=3cm, minimum height=2cm, align=center, fill=blue!10,rectangle, draw, rounded corners, text centered] {User \\ Verification \\ Process};
 \node (AccessDecision) [left=of UserVerification, draw, minimum width=3cm, minimum height=2cm, align=center, fill=blue!10,rectangle, draw, rounded corners, text centered] {Access \\Decision };

 \node (logging) [right=of UserVerification, draw, minimum width=3cm, minimum height=2cm, align=center, fill=blue!10,rectangle, draw, rounded corners, text centered] {Audit and\\Logging};

 \draw[arrow] (user) -- (request.west);
 \draw[arrow] (request.east) -- (data);
 \draw[arrow] (data.east) -- (device);
 \draw[arrow] (device) |- (policies);
 \draw[arrow] (policies) -| (assessment);
 \draw[arrow] (assessment.east) -- ++(.5cm, 0)|- (LowRisk.west);
 \draw[arrow] (assessment.east) -- (MediumRisk);
 \draw[arrow] (assessment.east) -- ++(.5cm, 0)|- (HighRisk.west);
 \draw[arrow] (LowRisk.east) -- (SimpleAuth);
 \draw[arrow] (MediumRisk.east) -- (MoreSteps);
 \draw[arrow] (HighRisk.east) -- (AdvancedAuth);
 \draw[arrow] (SimpleAuth) -| (UserVerification); 
 \draw[arrow] (MoreSteps) -| (UserVerification); 
 \draw[arrow] (AdvancedAuth) -| (UserVerification); 
 \draw[arrow] (UserVerification) -- (AccessDecision);

 \draw[arrow, dashed] (AccessDecision.west) -| (user.south);
 \draw[arrow] (UserVerification) -- (logging);

 \end{tikzpicture}}
 \caption{Risk Based Adaptive Authentication Workflow}
 \label{fig:user_transaction_flowchart}
\end{figure}
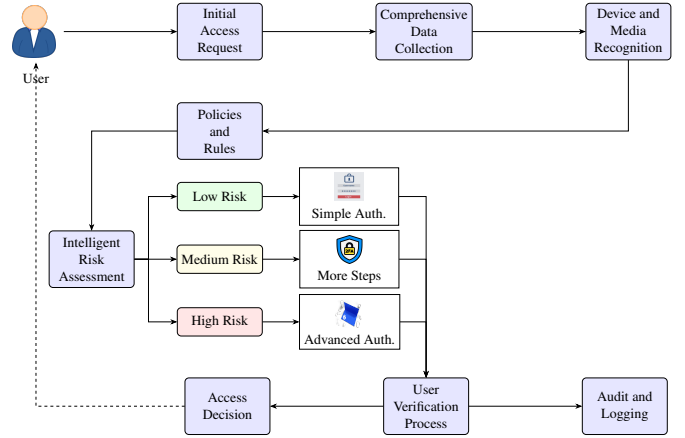

\gls{RBA} leverages \gls{ML} to dynamically assess the risk of login attempts by analyzing user features, including location, device type, time of day, and behavioral patterns~\cite{djosic2020machine}. This approach combines supervised, unsupervised, and reinforcement learning techniques to adapt security measures to evolving threats~\cite{hazratifard2022using}. For example, supervised algorithms such as logistic regression classify transactions, unsupervised methods like isolation forests detect anomalies, and reinforcement learning continuously optimizes risk policies. Through this multi-faceted analysis, \gls{RBA} enables effective, real-time risk assessments that help safeguard online transactions~\cite{burggraf2021beyond}.
  
Despite these benefits, implementing effective \gls{RBA} remains challenging due to privacy concerns and data diversity. Server-centric models~\cite{papaioannou2022toward, singh2022resilient, gupta2019driverauth} aggregate user data in centralized locations, raising risks of profiling and re-identification. Device-centric models~\cite{picard2023rlauth, progonov2022behavior} avoid centralization but face computational limitations, compatibility challenges, and high data requirements during the cold-start phase. A shared limitation across both paradigms is the \gls{NonIID} nature of user data, which complicates model training and often results in biased or inefficient learning.

 \gls{FL} offers a promising solution by enabling decentralized model training, which preserves user privacy and improves personalization. By keeping user data localized on devices, \gls{FL} ensures compliance with privacy regulations and reduces data breach risks. This approach also integrates individualized learning, enhancing the global model's accuracy and robustness by leveraging diverse user data without compromise.   However, implementing adaptive authentication within a \gls{FL} framework introduces a significant challenge: the inherently \gls{NonIID} nature of user features. While \gls{FL} enables decentralized model training with localized data to preserve privacy~\cite{kairouz2019advances, li2020federated}, the statistical heterogeneity across clients often causes local models to diverge~\cite{smith2017federated, fallah2020personalized}. As a result, naively aggregating these models can produce biased, unstable, or poorly generalized global models, ultimately degrading authentication accuracy and system robustness~\cite{liao2023hyperfed, LIANG2023110813}.

To address the challenges posed by \gls{NonIID} data in \gls{FL}, we propose \gls{FLRBA2}, a novel framework that transforms \gls{NonIID} user features into standardized similarity vectors. These vectors are generated by comparing registered and live features for each user before the learning process begins, creating inherently \gls{IID}-like representations suitable for federated aggregation. By incorporating these similarity vectors into \gls{FL}, \acrshort{FLRBA2} enables effective risk assessment while integrating individual user patterns into a robust global model, enhancing both security and usability. \acrshort{FLRBA2} leverages advanced techniques like \gls{DP} to protect sensitive information through controlled noise addition and \glspl{MAC} to ensure data integrity and tamper resistance. This comprehensive approach delivers a scalable, privacy-preserving solution for adaptive authentication that maintains high security standards without compromising system efficiency.

To the best of our knowledge, \acrshort{FLRBA2} is the first comprehensive framework for \gls{RBA} that unifies diverse feature categories across distributed users without centralizing raw data. At its core, the framework introduces a novel, mathematically proven similarity transformation that converts heterogeneous user features into standardized, \gls{IID}-like vectors. This transformation not only enables robust federated aggregation under \gls{NonIID} conditions but also preserves individual behavioral signatures for accurate and personalized risk assessment.

Compared to our previous work~\cite{fereidouni2024f}, \acrshort{FLRBA2} represents a fundamental advancement through three critical innovations that address privacy, security, and methodological limitations. First, the framework establishes a comprehensive \gls{RBA} feature taxonomy that seamlessly integrates an expanded set of contextual, knowledge-based, behavioral, biometric, and interaction-based features, employing mathematically rigorous similarity computation across all feature categories to ensure consistent and unbiased risk evaluation. Second, \acrshort{FLRBA2} employs a mathematically rigorous approach to mitigating feature-user correlation bias, enabling fair authentication decisions and ensuring unbiased \gls{FL} aggregation. Third, the framework provides formal security guarantees through game-based cryptographic proofs that demonstrate adaptive authentication security, privacy preservation via \gls{DP}, and federated aggregation correctness. With \gls{DP} inherently protecting against model inversion and inference attacks, \acrshort{FLRBA2} sets a new benchmark for scalable, privacy-preserving, and attack-resilient adaptive authentication in federated environments.

\subsection{Contributions}
This paper introduces \acrshort{FLRBA2}, a privacy-preserving federated framework for risk-based adaptive authentication, with the following core contributions:
\begin{itemize}
  \item \textbf{Unified Similarity-Based Transformation of \gls{NonIID} Data:} We propose a novel representation that converts heterogeneous \emph{knowledge-based, biometric, interaction-based, contextual, and behavioral} user features into per-session similarity vectors, enabling robust \gls{FL} over inherently \gls{NonIID} distributions.

 \item \textbf{Modality-Agnostic Risk Inference and Cold-Start Mitigation:} 
\acrshort{FLRBA2} supports clustering-based unsupervised risk labeling that feeds into lightweight supervised local models. 
This semi-supervised pipeline enables multi-tier adaptive authentication and effective handling of new or partial user sessions across diverse modalities.

  \item \textbf{Privacy-Preserving and Secure Federated Training:} We incorporate \gls{DP} and \glspl{MAC} to ensure privacy, integrity, and authenticity, formally proven through rigorous cryptographic analysis.

  \item \textbf{Comprehensive Real-World Evaluation:} We evaluate \acrshort{FLRBA2} on keystroke, mouse, and contextual datasets. Results demonstrate strong high-risk user detection with cluster-derived labels and provide a detailed privacy–utility trade-off analysis under differential privacy, validating the framework’s generalizability and robustness across modalities.
\end{itemize}

 \subsection{Organization}

The remainder of this paper is organized as follows. Section~\ref{sec:related_work} reviews the existing literature highlighting key research gaps that underscore the need for our work.  Section~\ref{sec:system_security_model} presents the system and security model, highlighting assumptions, threat models, and the security objectives essential for adaptive authentication. Section~\ref{sec:proposed_framework} introduces our proposed \acrshort{FLRBA2} framework, detailing its architecture, components, and key functionalities.  Section~\ref{sec:security_analysis} analyzes the security of the proposed \acrshort{FLRBA2} framework. Section~\ref{sec:performance_evaluation} evaluates the performance of \acrshort{FLRBA2} framework. Finally, Section~\ref{sec:conclusion} concludes the paper and discusses future research directions.

\section{Related Work} \label{sec:related_work}
This section reviews foundational research areas critical to privacy-preserving \gls{FL} for adaptive authentication. Specifically, we examine the privacy-security trade-offs in \gls{RBA} systems, \gls{PFL} techniques for \gls{NonIID} data, similarity evaluation for multimodal feature integration, and privacy-preserving authentication mechanisms. While advances exist in each domain, their holistic integration into a unified, scalable, privacy-aware federated framework remains largely unaddressed. These gaps directly motivate our \acrshort{FLRBA2} contribution.

\subsection{Risk-Based Authentication}

\gls{RBA} dynamically adjusts security measures based on the assessed risk of a login attempt, balancing security with user experience. By leveraging factors such as knowledge, biometrics, behavior, context, and tokens, \gls{RBA} systems evaluate the likelihood of fraudulent activity and apply appropriate authentication challenges.

Several studies have explored \gls{RBA} mechanisms and their integration with \gls{ML} techniques. Sepczuk and Kotulski~\cite{sepczuk2018new} propose a model incorporating contextual data, such as user security experience and service type, to refine risk assessment. Papaioannou et al.~\cite{papaioannou2022toward} focus on mobile authentication for border control, employing novelty detection to identify anomalies.  Picard and Pierre~\cite{picard2023rlauth} introduce RLAuth, a deep reinforcement learning-based system for adaptive challenge-response authentication. Singh et al.\cite{singh2022resilient} introduce a risk-based framework leveraging \gls{ML} to classify authentication contexts and adapt security measures accordingly. Wiefling et al.\cite{10.1145/3546069} collect and analyze large-scale real-world login data, applying ML-based optimization to enhance \gls{RBA} configuration and usability. Progonov et al.~\cite{progonov2022behavior} employ user behavior patterns for real-time authentication, achieving high accuracy and robustness against spoofing attacks.

Existing \gls{RBA} models can be broadly categorized into server-centric and device-centric approaches. Server-centric models, such as those in~\cite{sepczuk2018new, papaioannou2022toward, singh2022resilient, 10.1145/3546069,gupta2019driverauth}, aggregate user data in centralized servers for risk assessment. While this approach enables sophisticated \gls{ML} models and historical analysis, it introduces significant privacy concerns. Centralized data repositories are susceptible to data breaches, exposing sensitive user information to malicious actors. Moreover, the aggregation of user data facilitates the creation of detailed user profiles, enabling sophisticated profiling attacks and enabling inferences about individual user behavior and preferences. These risks extend beyond simple data breaches, as adversaries can exploit these centralized datasets to develop and deploy targeted attacks against specific individuals or groups.

In contrast, device-centric approaches, such as those explored in~\cite{picard2023rlauth, progonov2022behavior}, process authentication data locally, reducing external exposure. These models leverage behavioral biometrics, contextual risk factors, and reinforcement learning to assess authentication risks without centralized storage. 
However, they face computational constraints and limited data handling capabilities on resource-constrained devices. Cross-device compatibility presents additional challenges, as authentication profiles built on one device may not transfer seamlessly to another.  Additionally, they require significant data collection to establish reliable risk models, leading to cold-start challenges.

Despite these advancements, achieving a balance between security, privacy, and usability in \gls{RBA} remains an open challenge. Server-centric models introduce risks related to centralized data aggregation, while device-centric models struggle with cross-device functionality, cold-start problems, and computational limitations. These limitations highlight the need for alternative strategies that leverage distributed intelligence to address privacy concerns while maintaining the adaptability, efficiency, and cross-device compatibility of \gls{RBA} systems.
\subsection{Personalized Federated Learning}

\gls{FL} offers a promising alternative to both server-centric and device-centric \gls{RBA} approaches by avoiding centralized raw data storage while enabling cross-device model sharing through standardized similarity transformations. However, 
\gls{FL} faces substantial challenges under \gls{NonIID} data, a common real-world condition that often leads to convergence inefficiencies and degraded model performance. To address these limitations, \gls{PFL} approaches aim to adapt models to individual clients or client groups while retaining the collaborative advantages of \gls{FL}. These methods can be broadly categorized into four paradigms. \textit{Clustering-based approaches} group clients with statistically similar data distributions to enable more effective localized aggregation, as demonstrated by Liang et al.~\cite{LIANG2023110813} and Ghosh et al.~\cite{9832954}. \textit{Multi-task learning frameworks} model clients as distinct tasks with personalized objectives; foundational techniques were introduced by Smith et al.~\cite{smith2017federated}, while Fallah et al.~\cite{fallah2020personalized} proposed a meta-learning-based solution, and Kairouz et al.~\cite{kairouz2019advances} and Li et al.~\cite{li2020federated} provided comprehensive theoretical and system-level insights. \textit{Adaptive aggregation strategies}, such as FedDRL~\cite{nguyen2022fedDRL}, dynamically adjust client impact using deep reinforcement learning. \textit{Representation learning methods}, exemplified by HyperFed~\cite{liao2023hyperfed}, leverage hyperbolic geometry to capture hierarchical structures and address class imbalance in \gls{NonIID} environments.

However, existing methods inadequately address \gls{RBA}-specific requirements, including real-time processing constraints, data freshness, privacy-security trade-offs, and resilience to adversarial threats~\cite{9424138,lyu2020threats}. The temporal and behavioral variability of authentication contexts demands novel federated approaches that standardize heterogeneous user features for secure aggregation while preserving privacy.

\subsection{Similarity Evaluation}

Similarity evaluation is fundamental across authentication modalities, though its integration within \gls{FL} remains underexplored. Biometric authentication commonly employs measures such as \textit{cosine similarity} for face recognition~\cite{wang2015cosine}, while behavioral authentication often uses \textit{\gls{DTW}} for sequential data analysis~\cite{yao2021clustering}. Contextual authentication typically uses techniques like \textit{Jaccard similarity} for categorical data and \textit{Euclidean distance}  for continuous attributes, with more advanced methods such as \textit{Mahalanobis}~\cite{KADHIM2025100177} and \textit{Hausdorff distances}~\cite{232073} supporting high-dimensional comparisons.
However,  existing approaches focus primarily on single-modality similarity in centralized systems. The challenge of effectively combining diverse similarity measures across heterogeneous feature types in distributed \gls{FL} environments, particularly for \gls{NonIID} data aggregation, represents a significant research gap. Bridging this gap is essential for building robust, multimodal risk-based authentication systems.


\subsection{Privacy-Preserving Authentication}
Privacy-preserving authentication minimizes sensitive user information exposure but faces critical latency limitations for real-time scenarios. \glspl{ZKP}\cite{10367782,10902360} and anonymous credentials\cite{10636337} enable identity verification without data revelation but suffer from substantial proof generation overhead causing unacceptable authentication delays. Homomorphic encryption~\cite{REN2021105} operates orders of magnitude slower than plaintext operations and faces computational depth limitations that restrict complex \gls{ML} operations, compromising risk assessment precision. Secure \gls{MPC}\cite{10981699} enables collaborative decisions without exposing individual data but requires multiple communication rounds with latency scaling by participant count and complexity, making it unsuitable for latency-sensitive authentication. \gls{DP}\cite{LIU2025128653} offers computational efficiency through calibrated noise but creates accuracy-privacy trade-offs where strong privacy guarantees degrade model performance. Integrating these techniques into federated \gls{RBA} systems poses unique challenges, particularly for \gls{NonIID} behavioral and contextual data that may leak user-specific patterns under aggregation.

\subsection{Gaps in Existing Research}

While significant progress has been made in \gls{RBA}, \gls{PFL}, and privacy-preserving techniques, their intersection remains largely unexplored. Existing \gls{RBA} systems face a fundamental privacy-accuracy trade-off: server-centric approaches achieve high accuracy but compromise privacy through centralized data aggregation, while device-centric methods preserve privacy but suffer from limited cross-device compatibility and cold-start challenges. Current \gls{PFL} approaches, despite addressing \gls{NonIID} data challenges, inadequately meet \gls{RBA}'s stringent requirements for real-time processing, temporal data freshness, and adversarial resilience. Furthermore, existing similarity evaluation techniques operate primarily in centralized, single-modality contexts, lacking unified frameworks for heterogeneous feature aggregation in distributed \gls{FL} environments.
Our \acrshort{FLRBA2} framework directly addresses these convergent challenges by introducing a novel similarity-based transformation that converts \gls{NonIID} user features into \gls{IID} representations suitable for privacy-preserving federated aggregation, enabling robust, real-time risk assessment while maintaining user privacy across diverse authentication modalities.
\section{System and Security Model}\label{sec:system_security_model}
\begin{figure}[ht]\vspace{-0.2cm}
 \centering
 \resizebox{\linewidth}{!}{
 \begin{tikzpicture}[scale=0.9, every node/.style={scale=0.8}, node distance=3cm]
 
 \node (user) [draw=none, fill=none] {\includegraphics[width=1cm]{user.png}};
 \node [scale=0.7,draw,fill=blue!10, rounded corners,minimum height=5cm,minimum width=0.6\linewidth,below=0.2cm of user, align=center, xshift=0] (userbox) {
 \begin{minipage}{0.6\linewidth}User
 \begin{itemize}
 \item Submit Credentials,
 \item Interact with Client Application.\vspace{2.3cm}
 \end{itemize} 
 \end{minipage}};

 \node[align=center, right=of user, xshift=0.4cm] (client) {\includegraphics[width=1cm]{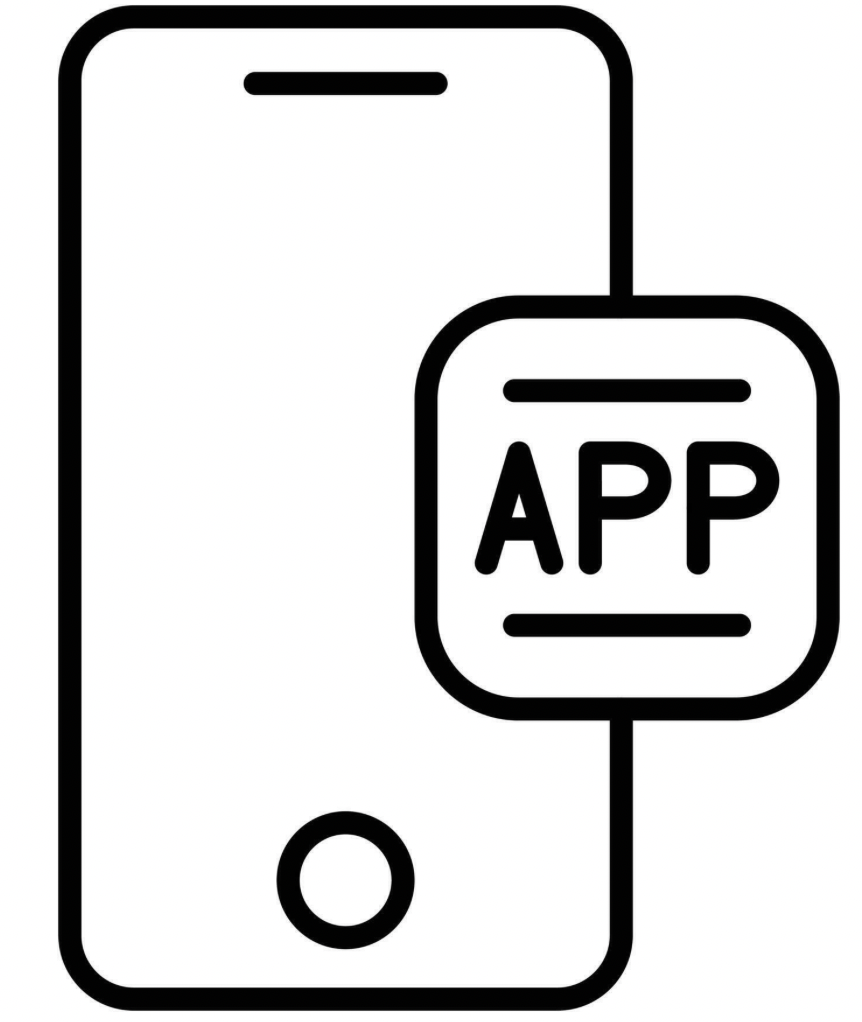}};
 \node [scale=0.7,draw,fill=blue!10, rounded corners,minimum height=5cm,minimum width=0.6\linewidth,below=0.22cm of client, align=center, xshift=0cm] (clientbox) {
 \begin{minipage}{0.6\linewidth}Client App
 \begin{itemize}
 \item Collect Data,
 \item Compute Similarity Vectors,
 \item Train Local Model,
 \item Evaluate Risk,
 \item Communicate with Auth. Server,
 \item Upload to and download from Cloud,
 \item Ensure Data Encryption,
 \item Apply MACs for Data Integrity.
 \end{itemize} 
 \end{minipage}};

 \node[align=center, right=of client] (server) {\includegraphics[width=1.5cm]{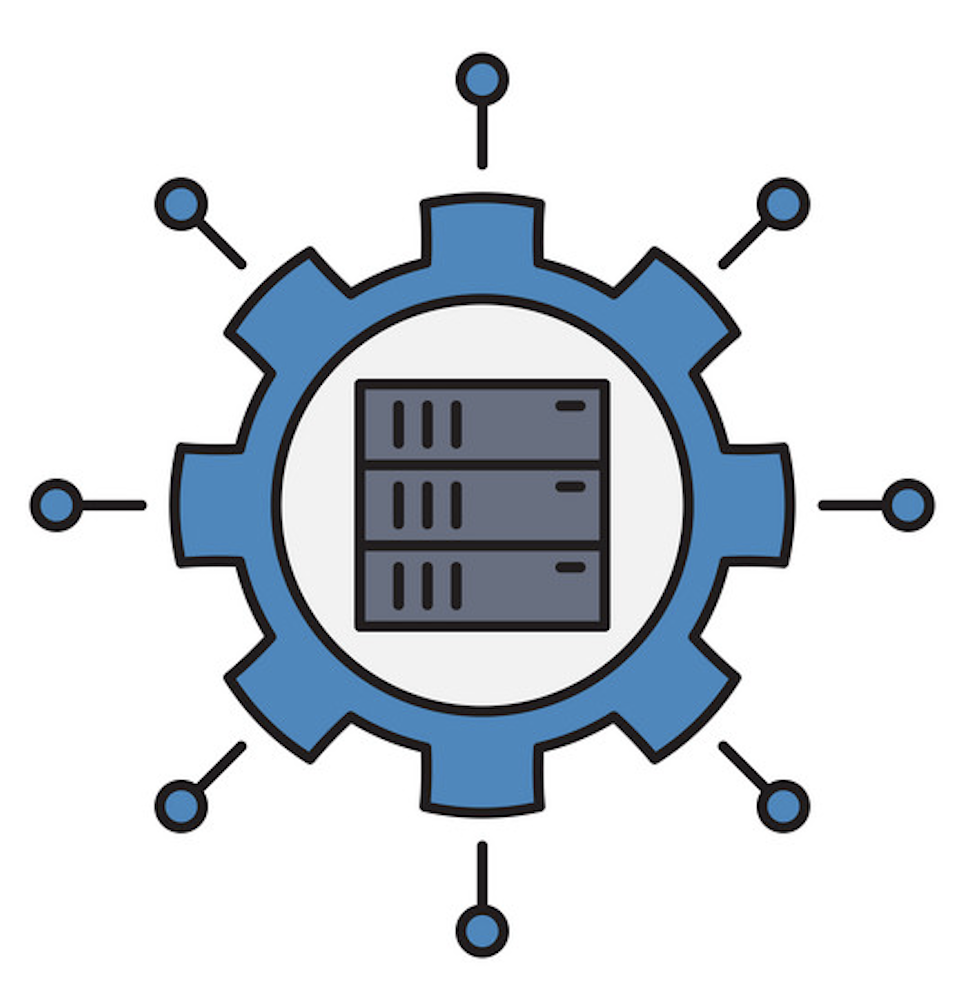}};
 \node [scale=0.7,draw,fill=blue!10, rounded corners,minimum height=5cm,minimum width=0.6\linewidth, below=0.07cm of server, align=center,xshift=0] (serverbox) {
 \begin{minipage}{0.6\linewidth}Server
 \begin{itemize}
 \item Select Auth. Methods,
 \item Authenticate Users,
 \item Aggregate Updates,
 \item Improve Global Model,
 \item Distribute Updated Global Model,
 \item Verify received MACs,
 \item Apply Privacy-Preserving Techniques.
 \end{itemize} 
 \end{minipage}};

 \node[draw, cloud, cloud puffs=8, cloud puff arc=120, aspect=2, fill=blue!20, minimum width=2cm, minimum height=1cm, above=of client,yshift=-2.4 cm] (cloud) {Cloud Storage};

 \draw[->] (user.east) -- ++(0.5,0) |- (client.west) node[midway,above,xshift=1.5cm] {Initial Access Request};

 \draw[->] ($(client.east) + (0,0.3)$) |- ($(server.west) + (0,0.3)$) node[midway,above,xshift=1.8cm] {Local Model};

 \draw[<-] ($(client.east) - (0,0.3)$) -- ($(server.west) - (0,0.3)$) node[midway,below] {Global Model};

 \draw[<->] (client) -- (cloud.south) node[midway, below, yshift=0.2cm, rotate=0]{Model Upload/Download};

[] \end{tikzpicture}}
 \caption{\acrshort{FLRBA2} System and Security Model}
 \label{fig:fl_model}
\end{figure}

Our proposed \acrshort{FLRBA2} framework comprises three primary components: the User, the Client Application, and the Authentication Server, as depicted in Figure \ref{fig:fl_model}.
The User initiates the authentication process by submitting credentials through the Client Application. The client application aggregates both explicit credentials (e.g., username, password) and implicit contextual data (e.g., access time, IP address, device type). It locally evaluates risk using a \gls{FL} model and transmits the computed risk level, along with the credentials, to the Authentication Server. The server selects appropriate authentication methods based on the received risk assessment and grants access accordingly.

\acrshort{FLRBA2} prioritizes user privacy by employing \gls{FL} and \gls{DP}. Model updates are based on similarity vectors, rather than raw user data, making it difficult to reconstruct original information. \gls{FL} keeps user data local, while the central server aggregates model updates without accessing raw data. 
\gls{DP} further enhances privacy by adding noise to model updates before transmission, rendering individual user data indistinguishable in the aggregated model. To ensure confidentiality of transmitted data, all communication between the client and server is encrypted. Time-stamped \glspl{MAC} are used to guarantee integrity and authenticity of messages, preventing tampering and replay attacks. By combining these robust security measures with \gls{FL} and privacy-preserving techniques, \acrshort{FLRBA2} provides a privacy-centric and secure framework for risk-based adaptive authentication.

The security of \acrshort{FLRBA2} is defined under a well-established threat model. 
The authentication server is assumed \textit{honest-but-curious}, executing the protocol faithfully but attempting passive inference, which is addressed through differential privacy. 
The end user is considered untrusted, while the client application that mediates authentication is assumed to be hardened against compromise or tampering. 
Communication channels are assumed secure, with encryption providing confidentiality against eavesdropping and time-stamped \glspl{MAC} guaranteeing message integrity and authenticity, thereby preventing tampering and replay by network adversaries.

\section{Proposed Approach}\label{sec:proposed_framework}
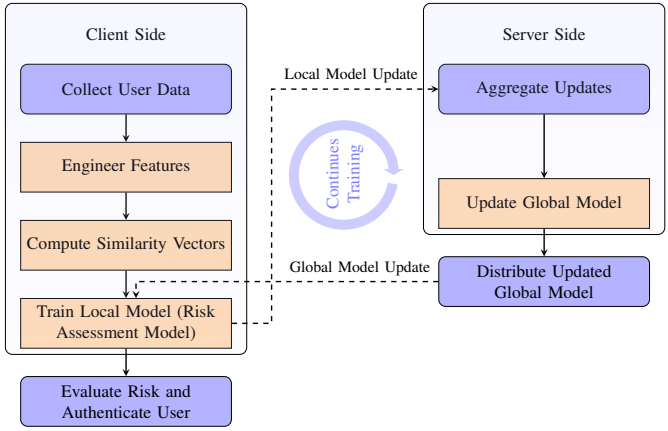
\begin{figure}[!htbp]
\centering
\resizebox{\linewidth}{!}{
\begin{tikzpicture}[node distance=2.1cm, title/.style={minimum height=1cm,minimum width=4.6cm,font = {\normalsize}, text centered}]
\tikzstyle{startstop} = [rectangle, rounded corners, minimum width=3cm, minimum height=1cm, align=justify, text centered, draw=black, fill=blue!30, text width=4cm]
\tikzstyle{process} = [rectangle, minimum width=3cm, minimum height=1cm, text centered, draw=black, fill=orange!30, text width=4cm, align=justify, text centered]
\tikzstyle{feedback} = [rectangle, minimum width=4cm, minimum height=3cm, text centered, text width=4cm]
\tikzstyle{arrow} = [thick,->,>=stealth]
\tikzstyle{data} = [rectangle, minimum width=4cm, minimum height=1cm, text centered, draw=black, fill=yellow!30, text width=4cm]

\node (cstep1) [startstop] {Collect User Data};
\node (cstep2) [process, below of=cstep1,yshift=0.53cm] {Engineer Features};
\node (cstep3) [process, below of=cstep2,yshift=0.53cm] {Compute Similarity Vectors};
\node (cstep4) [process, below of=cstep3,yshift=0.53cm] {Train Local Model (Risk
Assessment Model)};
\node (cstep5) [startstop, below of=cstep4,yshift=0.53cm] {Evaluate Risk and Authenticate User};

\draw [arrow] (cstep1) -- (cstep2);
\draw [arrow] (cstep2) -- (cstep3);
\draw [arrow] (cstep3) -- (cstep4);
\draw [arrow] (cstep4) -- (cstep5);

\node (cmodule) [title,above of=cstep1, yshift=-1cm] {Client Side};
\node (client-module)[top color=blue!30, bottom color=blue!10, rounded corners,minimum height=1cm,draw=black!100,fill opacity=0.1, fit={ (cstep1)(cstep2) (cstep3) (cstep4)(cmodule)}]{};

\node (sstep1) [startstop, right of=cstep1, xshift=6.3cm] {Aggregate Updates};
\node (sstep2) [process, below of=sstep1,yshift=-0.2cm] {Update Global Model};
\node (sstep3) [startstop, below of=sstep2,,yshift=0.53cm] {Distribute Updated Global Model};

\node (smodule) [title, right of=cmodule, xshift=6.3cm] {Server Side};

\node (server-module) [top color=blue!30, bottom color=blue!10, rounded corners,minimum height=1cm,draw=black!100,fill opacity=0.1,fit={ (sstep1) (sstep2) (smodule)}]{};

\draw [arrow] (sstep1) -- (sstep2);
\draw [arrow] (sstep2) -- (sstep3);

\node (feedback) [feedback, right of=cmodule, minimum height=2cm, text width=2cm, yshift=-2.8cm,xshift=2.3cm, text=blue!50,rotate=90] {{Continues Training}};

\draw [thick arrow,line width=0.2cm] ($(feedback.south)+(-0.05,-0.2)$) arc (350:0:1) ;

\draw [arrow,dashed] (cstep4.east) -- ++(0.8,0) |- (sstep1.west)node[midway, above, yshift=0cm, xshift=1.6cm] {\small{Local Model Update}};

\draw [arrow,dashed] (sstep3.west) -| ($(cstep4.north)+(0.2cm,0)$)node[midway, above, yshift=0cm, xshift=4.5cm] {\small{Global Model Update}};
\end{tikzpicture}}
\caption{\acrshort{FLRBA2} Methodology Process Flow}
\label{fig:methodology} \vspace{-0.4cm}
\end{figure}

Our proposed \acrshort{FLRBA2} framework addresses the challenges of \gls{NonIID} user features by transforming them into \gls{IID} similarity vectors, enabling robust \gls{FL} for enhanced security and privacy in adaptive authentication systems. As illustrated in Figure \ref{fig:methodology}, the framework consists of five key steps: \textit{(A) Feature Engineering}, which involves dynamically identifying, extracting, categorizing, and prioritizing user features---such as knowledge-based, biometric, behavioral, contextual, and interaction-based features---based on their relevance to risk assessment; 
\textit{(B) Similarity Evaluation}, which computes and normalizes similarity scores between registered and live features using feature-specific distance metrics;
\textit{(C) Similarity Vector Aggregation}, which creates a comprehensive similarity vector while maintaining approximate independence among features; \textit{(D) Local Risk Assessment Model}, which trains a local model on each device, leveraging the similarity vector to dynamically assess risk; and \textit{(E) Federated Learning Aggregation}, which integrates local models within a federated framework to enhance authentication strength while preserving user privacy. The subsequent sections provide a detailed exploration of each step.

\subsection{Feature Engineering}
Adaptive authentication leverages a diverse set of features to enhance security while ensuring a seamless user experience.  These features fall into five main categories:  (1) knowledge-based, (2) biometric, (3) behavioral, (4) contextual, and (5) interaction-based features. 
Knowledge-based features include passwords and security questions. These features rely on information the user knows, making them a traditional yet essential part of the authentication process. Biometric features encompass fingerprint recognition, facial recognition, iris recognition, and voice recognition. These features use unique biological characteristics of the user, providing a higher level of security due to their uniqueness and difficulty to replicate.
Behavioral features consist of typing patterns, mouse movements, and gait analysis. These features analyze the user's behavior and interaction with devices, adding an additional layer of security by detecting anomalies in habitual actions. Contextual features involve location (GPS data), time of access, the device used (device fingerprinting), and IP address. These features provide context to the authentication attempt, ensuring that it aligns with the user's usual patterns and environment.
Lastly, interaction-based features include application usage patterns, browser history, and social network activity. These features monitor the user's interactions with applications and online services, contributing to a comprehensive understanding of the user's behavior and enhancing the system's ability to detect unauthorized access attempts.

\begin{figure}[!htbp]
 \begin{center}
 \resizebox{\linewidth}{!}{
 \begin{tikzpicture}[line width=0.035cm]
 \large
 \node (main) [draw,fill=blue!10, rounded corners,minimum height=1cm,minimum width=4cm] {Adaptive Auth. Features};
 
 \node (knowledge) [draw,fill=blue!10, rounded corners,minimum height=1cm,minimum width=4cm,above right of=main,xshift=4.4cm,yshift=8.4cm] {Knowledge-Based};
 \node (knowledge1) [draw,fill=blue!10, rounded corners,minimum height=1cm,minimum width=7cm,above right of=knowledge,xshift=6cm,yshift=0cm] {Passwords};
 \node (knowledge2) [draw,fill=blue!10, rounded corners,minimum height=1cm,minimum width=7cm,below of=knowledge1,yshift=-0.4cm] {Security Questions};

 \node (biometric) [draw,fill=blue!10, rounded corners,minimum height=1cm,minimum width=4cm, below of=knowledge,yshift=-3.2cm] {Biometric}; 
 \node (biometric1) [draw,fill=blue!10, rounded corners,minimum height=1cm,minimum width=7cm,above right of=biometric,xshift=6cm,yshift=1.4cm] {Fingerprint Recognition};
 \node (biometric2) [draw,fill=blue!10, rounded corners,minimum height=1cm,minimum width=7cm,below of=biometric1,yshift=-0.4cm] {Facial Recognition};
 \node (biometric3) [draw,fill=blue!10, rounded corners,minimum height=1cm,minimum width=7cm,below of=biometric2,yshift=-0.4cm] {Iris Recognition};
 \node (biometric4) [draw,fill=blue!10, rounded corners,minimum height=1cm,minimum width=7cm,below of=biometric3,yshift=-0.4cm] {Voice Recognition};

 \node (behavioral) [draw,fill=blue!10, rounded corners,minimum height=1cm,minimum width=4cm, below of=biometric,yshift=-3.9cm] {Behavioral};
 \node (behavioral1) [draw,fill=blue!10, rounded corners,minimum height=1cm,minimum width=7cm,above right of=behavioral,xshift=6cm,yshift=0.7cm] {Typing Patterns};
 \node (behavioral2) [draw,fill=blue!10, rounded corners,minimum height=1cm,minimum width=7cm,below of=behavioral1,yshift=-0.4cm] {Mouse Movements};
 \node (behavioral3) [draw,fill=blue!10, rounded corners,minimum height=1cm,minimum width=7cm,below of=behavioral2,yshift=-0.4cm] {Gait Analysis};

 \node (contextual) [draw,fill=blue!10, rounded corners,minimum height=1cm,minimum width=4cm, below of=behavioral,yshift=-3.8cm] {Contextual};
 \node (contextual1) [draw,fill=blue!10, rounded corners,minimum height=1cm,minimum width=7cm,above right of=contextual,xshift=6cm,yshift=1.4cm] {Device Used (Device Fingerprinting)};
 \node (contextual2) [draw,fill=blue!10, rounded corners,minimum height=1cm,minimum width=7cm,below of=contextual1,yshift=-0.4cm] {Location (GPS data)};
 \node (contextual3) [draw,fill=blue!10, rounded corners,minimum height=1cm,minimum width=7cm,below of=contextual2,yshift=-0.4cm] {IP Address};
 \node (contextual4) [draw,fill=blue!10, rounded corners,minimum height=1cm,minimum width=7cm,below of=contextual3,yshift=-0.4cm] {Time of Access};

 \node (interaction) [draw,fill=blue!10, rounded corners,minimum height=1cm,minimum width=4cm, below of=contextual,yshift=-3.8cm] {Interaction-Based};
 \node (interaction1) [draw,fill=blue!10, rounded corners,minimum height=1cm,minimum width=7cm,above right of=interaction,xshift=6cm,yshift=0.7cm] {Application Usage Patterns};
 \node (interaction2) [draw,fill=blue!10, rounded corners,minimum height=1cm,minimum width=7cm,below of=interaction1,yshift=-0.4cm] {Browser History};
 \node (interaction3) [draw,fill=blue!10, rounded corners,minimum height=1cm,minimum width=7cm,below of=interaction2,yshift=-0.4cm] {Social Network Activity};

 \draw[-latex] (main) |- (knowledge);
 \draw[-latex] (knowledge) |- (knowledge1);
 \draw[-latex] (knowledge) |- (knowledge2);
 
 \draw[-latex] (main) |- (biometric);
 \draw[-latex] (biometric) |- (biometric1);
 \draw[-latex] (biometric) |- (biometric2);
 \draw[-latex] (biometric) |- (biometric3);
 \draw[-latex] (biometric) |- (biometric4);
 
 \draw[-latex] (main) -- (behavioral);
 \draw[-latex] (behavioral) |- (behavioral1);
 \draw[-latex] (behavioral) -- (behavioral2);
 \draw[-latex] (behavioral) |- (behavioral3);
 
 \draw[-latex] (main) |- (contextual);
 \draw[-latex] (contextual) |- (contextual1);
 \draw[-latex] (contextual) |- (contextual2);
 \draw[-latex] (contextual) |- (contextual3);
 \draw[-latex] (contextual) |- (contextual4);
 
 \draw[-latex] (main) |- (interaction);
 \draw[-latex] (interaction) |- (interaction1);
 \draw[-latex] (interaction) -- (interaction2);
 \draw[-latex] (interaction) |- (interaction3);
 
 \end{tikzpicture}
 }
 \caption{Categorization of Features for Adaptive Authentication}
 \label{fig:adaptive-auth-features}
 \end{center}
 \vspace{-0.5cm}
\end{figure}
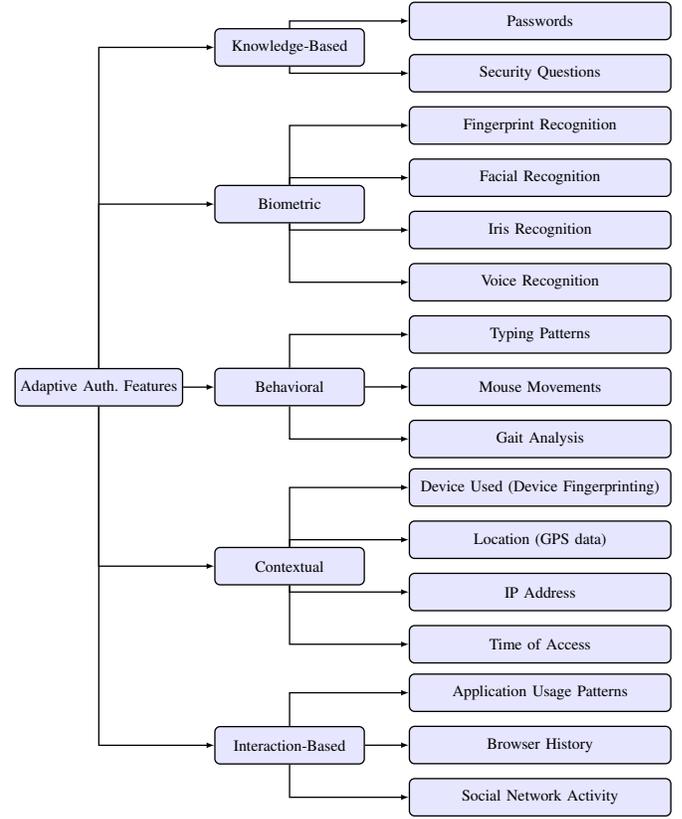

\subsection{Similarity Evaluation: A Technique to Aggregate \gls{NonIID} Features}\label{sec:similarity_computation}
This section proposes a method for aggregating features that are not \gls{IID} and cannot be easily generalized. The method involves categorizing features and measuring the similarity between registered and live user data, or using alternative methods when registered data is not available. It transforms user-specific features into similarity vectors, leveraging distance computation methodologies. This ensures the effective integration of features before starting the learning process for each user, and subsequently integrates each user's learning within a \gls{FL} framework into a general model.

To measure the similarity\textit{ \( \text{Sim} \)} between reference features \( F_{\text{Ref}} \) (e.g., registered or historical features) and live features \( F_{\text{Live}} \), and to create a similarity vector that combines all these features, we evaluate the distance computation for each feature category.

\subsubsection{Knowledge-Based Features}

For binary, knowledge-based features \textit{F} such as passwords and security questions, similarity is evaluated as either a match or not. Here, \( F_{\text{Ref}} \) denotes the registered (reference) feature. A perfect match yields a similarity score of 1, while any discrepancy results in a score of 0.
\begin{align*}
\text{Sim}_{\textit{F}} = 
\begin{cases} 
1 & \text{if } F_{\text{Ref}} = F_{\text{Live}} \\
0 & \text{otherwise}
\end{cases}
\end{align*}
This binary approach provides a straightforward similarity evaluation for each knowledge-based feature, such as passwords or security questions.
\subsubsection{Biometric Features}

To measure the similarity between biometric features \textit{F} (e.g., fingerprint, facial, iris, or voice recognition), a correlation function like cosine similarity is often used. Cosine similarity evaluates the cosine of the angle between two vectors, where a higher similarity score indicates a closer match between the biometric data.
\begin{align*}
\text{Sim}_{\textit{F}} = \text{Cosine}(F_{\text{Ref}}, F_{\text{Live}})
\end{align*}
where \( F_{\text{Ref}} \) represents the reference data (e.g., registered feature), \( F_{\text{Live}} \) represents the live data point, and cosine similarity is defined as:
\begin{align*}
\text{Cosine}(F_{\text{Ref}}, F_{\text{Live}}) = \frac{F_{\text{Ref}} \cdot F_{\text{Live}}}{\|F_{\text{Ref}}\| \|F_{\text{Live}}\|}
\end{align*}
Biometric features are generally represented as high-dimensional vectors, and cosine similarity effectively measures the angle between these vectors. This approach provides a robust method for comparing biometric data, which often involves complex patterns and high-dimensional representations.

\subsubsection{Behavioral Features}

\glsentryfull{DTW} is a powerful technique for measuring the similarity between two time series sequences, such as typing patterns, mouse movements, or touchscreen gestures. It is particularly well-suited for analyzing human behavior, as it accounts for variations in timing and sequence length. 
The \gls{DTW} distance between two sequences, \( F_{\text{Ref}} = \{ r_1, \ldots, r_m \} \) and \( F_{\text{Live}} = \{ l_1, \ldots, l_n \} \), minimizes cumulative distance while allowing for non-linear alignments in time. This is achieved by first computing a cost matrix \( C \), where each element \( C(i, j) \) represents the distance (e.g., Euclidean) between \( r_i \) and \( l_j \). An accumulated cost matrix \( D \) is then calculated iteratively using the relation:
\[
D(i, j) = C(i, j) + \min(D(i-1, j), D(i, j-1), D(i-1, j-1)).
\]

The \gls{DTW} distance is obtained as \( D(m, n) \), the value at the bottom-right of \( D \), with the alignment path recoverable through backtracking. \gls{DTW} is widely used in time-series analysis due to its ability to handle temporal variations effectively.

To compute the similarity between a reference behavioral pattern (\( F_{\text{Ref}} \)) and a live behavioral pattern (\( F_{\text{Live}} \)), we use the formula:
\begin{align*}
\text{Sim}_{\textit{F}} = 1 - \frac{\text{DTW Distance}(F_{\text{Ref}}, F_{\text{Live}})}{\text{Max DTW Distance}}.
\end{align*}
Here, \(\text{Max DTW Distance}\) is a normalization constant derived from the maximum observed \gls{DTW} distance for the same feature across a user's historical data. This normalization ensures similarity scores are scaled between 0 and 1, maintaining consistency and mitigating sensitivity  to minor behavioral variations.

When insufficient registered data is available as direct reference patterns \( F_{\text{Ref}} \), 
a centroid-based approach can be applied using \gls{DTW} Barycenter Averaging (DBA) to construct a representative user profile that captures the user’s typical behavioral pattern.  Unlike simple averaging, the \gls{DTW} 
DBA method is used to compute a centroid that aligns sequences temporally, ensuring that variations in timing are accurately captured. The \gls{DTW} barycenter 
\( \text{Cen}_{\textit{F}} \) minimizes the average \gls{DTW} distance to all historical behavior sequences, resulting in a representative pattern that respects sequence alignment and timing. The live data \( F_{\text{Live}} \) is then compared to this DTW-aligned centroid using \gls{DTW} or other distance metrics, allowing for accurate similarity measurement even with limited registered data.
By leveraging DTW, distance-based inverse scaling, and centroid-based methods, this approach effectively measures the similarity between behavioral patterns, providing a robust framework for user authentication and behavior analysis.

{While} \gls{DTW} {provides optimal sequence alignment, its $\mathcal{O}(N^2)$ time and space complexity limits its practicality in real-time or edge-deployed applications. To address this, we adopt} \text{FastDTW}, a linear-time approximation that leverages multilevel resolution projection and locally constrained warping~\cite{10909374}. This yields scalable alignment with accuracy comparable to DTW, making it well-suited for on-device behavioral authentication.




\subsubsection{Contextual Features}

Contextual features are essential for assessing the risk of login attempts, including factors such as the device used, the user's location, IP address, and access time. Deviations from a user's typical behavior in these areas may signal potential fraud. We categorize contextual feature similarity \(\text{Sim}\) into two types: set membership and distance-based metrics.

For the \textit{Device Used} feature, similarity is assessed by determining if the live device is part of the set of devices previously used by the user. This similarity, \( \text{Sim}_{\textit{F}} \), is evaluated with a set-based membership function:
\begin{align*}
\text{Sim}_{\textit{F}} = 
\begin{cases} 
1 & \text{if } F_{\text{Live}} \in F_{\text{Ref}} \\
0 & \text{otherwise}
\end{cases}
\end{align*}
where \( F_{\text{Ref}} \) denotes the set of all devices previously used by the user, and \( F_{\text{Live}} \) represents the current device. This set-based approach provides a more accurate measure of similarity than a simple binary match.

For the second type of contextual features, including location, IP address and time of access, similarity is calculated using a general formula:
\begin{align*}
\text{Sim}_{\textit{F}} = 1 - \frac{\text{Distance}_{\textit{F}}(F_{\text{Ref}}, F_{\text{Live}})}{\text{Max Distance}_{\textit{F}}}.
\end{align*}
where \( F_{\text{Ref}} \) represents the reference data (e.g., historical location centroid), \( F_{\text{Live}} \) represents the live data point (e.g., current login location), \( \text{Distance}_{\textit{F}} \) is a feature-specific distance metric, and \( \text{Max Distance}_{\textit{F}} \) is the maximum possible distance for the feature. Here, \( \text{Max Distance}_{\textit{F}} \) serves as a normalization constant, ensuring that the similarity score is bounded between 0 and 1 and decreases gradually as deviations increase. This approach accommodates natural contextual shifts while identifying significant anomalies that may indicate unauthorized access.

For location data, the distance metric \( \text{Distance}_{\text{Location}} \) is computed using the Haversine formula:\\ \\
\resizebox{\linewidth}{!}{$
\text{Distance}_{\text{Location}} = 2R \cdot \arcsin\left(\sqrt{\sin^2\left(\frac{\Delta \phi}{2}\right) + \cos(\phi_1) \cos(\phi_2) \sin^2\left(\frac{\Delta \lambda}{2}\right)}\right)
$}.\\ \\
Here \( R \) is the Earth's radius, \( \Delta \phi = \phi_2 - \phi_1 \) is the difference in latitude, and \( \Delta \lambda = \lambda_2 - \lambda_1 \) is the difference in longitude. The latitudes (\( \phi_1, \phi_2 \)) and longitudes (\( \lambda_1, \lambda_2 \)) must be expressed in radians for accurate calculations. If registered location data is unavailable, a \textit{centroid} derived from historical latitude and longitude coordinates can be used as a representative location. This centroid reflects the typical location from historical data, serving as a reference for comparisons.

For software versions (e.g., Browser Version, OS Version), which are numeric but may not follow semantic versioning rigorously, we define version drift similarity as:
\begin{align*}
\text{Sim}_{\text{Version}} = 1 - \frac{|v_{\text{Live}} - v_{\text{Ref}}|}{\max(v_{\text{Live}}, v_{\text{Ref}})}
\end{align*}
This formulation provides a normalized measure of deviation between observed and reference version values, penalizing substantial mismatches while allowing minor version updates.

\subsubsection{Interaction-Based Features}

Similarity for interaction-based features (e.g., application usage patterns, browser history, social network activity) is evaluated using Jaccard similarity for categorical data. Jaccard similarity measures the similarity between finite sets, defined as the size of the intersection divided by the size of the union of the sets.
\begin{align*}
\text{Sim}_{\textit{F}} = \frac{|F_{\text{Ref}} \cap F_{\text{Live}}|}{|F_{\text{Ref}} \cup F_{\text{Live}}|}
\end{align*}

Interaction-based features are often categorical, representing sets of actions or preferences. Jaccard similarity is well-suited for comparing such categorical data, as it measures the overlap between sets relative to their union size, providing an intuitive and effective similarity score for these feature types.
If registered interaction data \(F_{\text{Ref}}\) is unavailable, population baselines can be established. Live data \(F_{\text{Live}}\) is then compared to these baselines using similar distance measures.

\begin{enumerate}[topsep=1ex, itemsep=1ex, wide, font=\itshape, labelwidth=!, labelindent=0pt, label*=B.\arabic*.]
\item \textit{Adaptive Similarity Evaluation with Decay Factor for Temporal Patterns:}
To account for historical influence, especially for features that show temporal stability with evolving patterns (such as location, IP address, or time of access), we introduce a decay factor \( \alpha \) into the similarity evaluation. This factor balances the weight between historical and current data, allowing the model to adapt to evolving user behavior while preserving the stability of long-term trends.
For features that rely on historical centroids, we calculate an updated centroid \(\text{Cen}_{\textit{F}}^{\text{new}} \) as follows:
\begin{align*}
\text{Cen}_{\textit{F}}^{\text{new}} = \alpha \cdot \text{Cen}_{\textit{F}}^{\text{hist}} + (1 - \alpha) \cdot F_{\text{Live}},
\end{align*}
where \(\text{Cen}_{\textit{F}}^{\text{hist}} \) is the historical centroid, representing the accumulated central tendency of past values, \( F_{\text{Live}} \) denotes the current feature value, based on the latest user data, and  \( \alpha \) is the decay factor, balancing the weight of historical versus current data. A smaller \( \alpha \) emphasizes recent user behavior, while a larger \( \alpha \) maintains a stronger historical perspective.
The decay factor \( \alpha \) can be adjusted based on the volatility and significance of each feature. For instance, a smaller \( \alpha \) might be used for features like IP address, which can change frequently, while a larger \( \alpha \) might be used for features like device type, which tend to be more stable. By incorporating this centroid-based decay factor, the \acrshort{FLRBA2} framework effectively balances historical influence and real-time adaptation, leading to more accurate and robust similarity evaluations and risk assessments.

\item \textit{Aggregating Similarities:}
Feature-specific transformations independently convert each raw feature into a normalized similarity score. 
These transformations reduce correlations within the similarity vector, achieving approximate independence for robust aggregation. As a result, the final similarity vector \(\bm{Sim} \) is constructed as:
\begin{align*}
\bm{Sim} = (\text{Sim}_{\textit{F}_1}, \text{Sim}_{\textit{F}_2}, \dots, \text{Sim}_{\textit{F}_n})
\end{align*}
By evaluating similarity vectors between registered and live user features, we can assess how closely a user's current profile aligns with their historical behavior, allowing for a comprehensive, personalized risk assessment. This approach leverages privacy-preserving and precise distance computations to maintain user confidentiality. When registered features are unavailable, alternative techniques, such as those based on historical patterns and population data, ensure robust and adaptable authentication, enhancing accuracy even in cases with limited prior data.
\begin{thm}[\textbf{Bounded Average Cross-Group Correlation via Similarity Transformation}]
\label{thm:bounded_correlation}
Let feature similarities $\bm{Sim} = (Sim_{F_1}, \ldots, Sim_{F_n})$ be partitioned into $m$ semantic groups 
$G_1, \ldots, G_m$ (e.g., geographic, device, temporal, behavioral). 
Assume each similarity variable is non-degenerate and standardized to zero mean and unit variance, and that 
the average cross-group correlation is bounded:
\[
\max_{a \neq b} \; \frac{1}{|G_a| \cdot |G_b|} 
\sum_{i \in G_a,\, j \in G_b} 
  \left| \mathrm{Corr}(Sim_{F_i}, Sim_{F_j}) \right| \;\leq\; \epsilon_{\text{avg}} \ll 1.
\]
Then there exists an absolute constant $C>0$ such that the total variation distance satisfies
\[
\| \mathbb{P}_{\bm{Sim}} - \prod_{j=1}^n \nu_j \|_{\mathrm{TV}} 
\;\leq\; C \cdot \frac{m(m-1)}{2}\, \epsilon_{\text{avg}},
\]
where $\mathbb{P}_{\bm{Sim}}$ denotes the joint distribution of the feature similarity vector $\bm{Sim}$, 
$\nu_j$ is the marginal distribution of $Sim_{F_j}$, and 
$\|\cdot\|_{\mathrm{TV}}$ denotes total variation distance, 
$\|P-Q\|_{\mathrm{TV}} = \sup_{A}|P(A)-Q(A)|$.
\end{thm}

\begin{proof}
The semantic partitioning allows strong dependencies within groups (e.g., correlations among device attributes) 
but requires cross-group dependencies to be weak on average. 
Let $X_i = Sim_{F_i}$; standardization implies
\[
\rho_{ij} \;=\; \big|\mathbb{E}[X_iX_j] - \mathbb{E}[X_i]\mathbb{E}[X_j]\big| 
\;=\; \big|\mathrm{Corr}(X_i, X_j)\big|.
\]

Decompose pairwise dependence:
\[
\sum_{i<j} \rho_{ij} \;=\; 
  \sum_{k=1}^m \sum_{\substack{i,j \in G_k\\ i<j}} \rho_{ij}
  \;+\; \sum_{a<b} \sum_{i \in G_a,\, j \in G_b} \rho_{ij}.
\]
The first (intra-group) term is unconstrained; the second (cross-group) term is controlled by the assumption:
\[
\frac{1}{|G_a|\cdot |G_b|} \sum_{i \in G_a,\, j \in G_b} \rho_{ij}
  \;\leq\; \epsilon_{\text{avg}} \qquad (\forall a \neq b).
\]
Summing over all $\binom{m}{2}$ cross-group pairs yields
\[
\sum_{a<b} \frac{1}{|G_a|\cdot |G_b|} 
       \sum_{i \in G_a,\, j \in G_b} \rho_{ij}
   \;\leq\; \binom{m}{2}\,\epsilon_{\text{avg}}.
\]

A group-structured extension of the Diaconis--Freedman inequality 
(see~\cite{diaconis1980finite}) bounds the total variation between the joint distribution and the product of marginals by this averaged cross-block dependence:
\[
\| \mathbb{P}_{\bm{Sim}} - \prod_{j=1}^n \nu_j \|_{\mathrm{TV}}
  \;\leq\; C \sum_{a<b} 
      \frac{1}{|G_a|\cdot |G_b|} 
      \sum_{i \in G_a,\, j \in G_b} \rho_{ij}.
\]
Here the factor $1/(|G_a|\cdot|G_b|)$ exactly cancels the $|G_a||G_b|$ multiplicity of terms inside each double sum, so each cross-group block contributes at most $\epsilon_{\text{avg}}$. 
Since there are $\binom{m}{2}$ such blocks, we obtain
\[
\| \mathbb{P}_{\bm{Sim}} - \prod_{j=1}^n \nu_j \|_{\mathrm{TV}}
  \;\leq\; C \cdot \binom{m}{2}\,\epsilon_{\text{avg}}.\qedhere
\]
\end{proof}

\begin{thm}[\textbf{Group-Structured Indistinguishability}]
\label{thm:group_indistinguishability}
For similarity vectors from Theorem~\ref{thm:bounded_correlation}, any PPT adversary $\mathcal{A}$ satisfies:
\begin{align*}
\text{Adv}^{\text{indep}}_{\mathcal{A}}(\lambda) :=& \left| \Pr[\mathcal{A}(\bm{Sim}_{1:K}) = 1] - \Pr[\mathcal{A}(\bm{Sim}'_{1:K}) = 1] \right| \\
 \leq & K \cdot \frac{m(m-1)}{2} \epsilon_{\text{avg}}
\end{align*}
where $\bm{Sim}'_i \sim \prod_{j=1}^n \nu_j$ are independent samples.
\end{thm}
\begin{proof}
From Theorem~\ref{thm:bounded_correlation}, we have the single-user total variation bound:
\[
\| \mathbb{P}_{\bm{Sim}} - \prod_{j=1}^n \nu_j \|_{\mathrm{TV}} \leq \frac{m(m-1)}{2} \epsilon_{\text{avg}}
\]

The adversary's distinguishing advantage is bounded by the total variation distance between the two $K$-user distributions. By the coupling lemma~\cite{lindvall2002lectures}, for any two probability distributions $P$ and $Q$, the distinguishing advantage of any adversary $\mathcal{A}$ satisfies:
\[
\left| \Pr_{X \sim P}[\mathcal{A}(X) = 1] - \Pr_{Y \sim Q}[\mathcal{A}(Y) = 1] \right| \leq \|P - Q\|_{\text{TV}}
\]

Since similarity vectors are sampled independently across users in both games, the $K$-user distributions are:
\[
P_{\text{real}} = \mathbb{P}_{\bm{Sim}}^{\otimes K}, \quad P_{\text{ideal}} = \left(\prod_{j=1}^n \nu_j\right)^{\otimes K}
\]

By the tensorization (subadditivity) property of total variation distance, for independent samples:
\[
\| P^{\otimes K} - Q^{\otimes K} \|_{\text{TV}} \leq K \cdot \|P - Q\|_{\text{TV}}
\]

Applying this to our distributions:
\[
\| \mathbb{P}_{\bm{Sim}}^{\otimes K} - \left(\prod_{j=1}^n \nu_j\right)^{\otimes K} \|_{\text{TV}} \leq K \cdot \frac{m(m-1)}{2} \epsilon_{\text{avg}}
\]

Therefore, the adversarial distinguishing advantage is bounded by $K \cdot \frac{m(m-1)}{2} \epsilon_{\text{avg}}$.
\end{proof}

\begin{remark}[\textbf{Implementation Justification}]
\label{rem:implementation}
Although Theorem~\ref{thm:bounded_correlation} acknowledges that intra-group correlations may persist, the FL-RBA$^2$ framework operates directly on per-feature similarity vectors rather than aggregating at the group level. This design is justified by three considerations: (1) the similarity transformation of raw features reduces the magnitude of cross-group dependencies in practice; (2) while Gaussian noise applied to local model updates during federated aggregation does not strictly remove feature-level correlation, it reduces the influence of any residual dependence on the aggregated global model, thereby limiting adversarial exploitation; and (3) federated averaging across many users naturally dampens systematic correlation effects, as residual user-specific dependencies tend to cancel out, improving the robustness of convergence. Empirical evaluation across modalities confirms that classification performance is preserved under this design, supporting the practical validity of the semantic independence assumption.
\end{remark}

\vspace{-0.5cm}
\end{enumerate}
\subsection{Local Risk Assessment Model}
The \acrshort{FLRBA2} framework employs a decentralized Local Risk Assessment Model on each user device to evaluate real-time authentication risk. This model analyzes user behavior and contextual data, enhancing privacy by retaining sensitive data locally and reducing reliance on centralized storage.

\begin{enumerate}[topsep=1ex, itemsep=1ex, wide, font=\itshape, labelwidth=!, labelindent=0pt, label*=C.\arabic*.]
\item \textit{Dynamic Risk Scoring and Similarity-Based Evaluation:}
Each device independently collects and processes user-specific data, including interaction patterns, environmental context, and device usage metrics. These data are transformed into similarity vectors, where each component quantifies the resemblance between reference (historical or registered) and current (live) user states across predefined features. The resulting similarity vector $\bm{Sim}_i = (\text{Sim}_{\textit{F}_1}, \text{Sim}_{\textit{F}_2}, \dots, \text{Sim}_{\textit{F}_n})$ for user $\mathcal{U}_i$ serves as the input to a local \gls{ML} model that dynamically evaluates authentication risk.
Formally, the local risk score $\text{Risk}_{\mathcal{U}_i}$ is computed as:
\begin{align*}
    \text{Risk}_{\mathcal{U}_i} = \mathcal{M}_{\theta}(\bm{Sim}_i),
\end{align*}
where $\mathcal{M}_{\theta}$ denotes a \gls{ML} model (e.g., logistic regression, decision tree, or neural network) trained on similarity vectors with parameters $\theta$. This model captures complex, potentially non-linear relationships among features and learns to discriminate between legitimate and anomalous behavior. Feature relevance is implicitly captured through the model’s learned parameters, enabling adaptive and context-aware risk assessment across diverse modalities.

\item \textit{Risk Categorization and Thresholding Strategy:} \label{sec:risk_labeling}
\acrshort{FLRBA2} employs a semi-supervised pipeline for risk inference: unsupervised K-means clustering generates pseudo-labels from per-session similarity representations (\(k=3\) for low/medium/high), followed by supervised L2-regularized logistic regression trained on these cluster assignments; models are later aggregated via federated aggregation. This yields interpretable multi-level classification without requiring ground-truth labels.
Because per-feature scores are defined as \emph{similarities} in \([0,1]\) (larger = more similar), legitimate sessions (lower risk) tend to have \emph{larger} coordinates and thus centroids \emph{farther} from the origin in similarity space, whereas anomalous sessions (higher risk) have smaller norms. We therefore assign risk by \emph{inverse} centroid-norm ranking:
\[
  \text{Risk}(C_i) = \text{rank}\!\big(-\|\boldsymbol{\mu}_i\|_2\big),
\]
so clusters with smaller \(\|\boldsymbol{\mu}_i\|_2\) correspond to higher risk categories.

\item \textit{Adaptive Decision-Making and Cold-Start Optimization:}
Cold-start scenarios present fundamental challenges when new users lack sufficient historical data for reliable behavioral modeling and similarity computation. 
\acrshort{FLRBA2} addresses this through a conservative-to-adaptive thresholding strategy combined with federated aggregation, allowing newly enrolled users to benefit from collective authentication insights embedded in the global model without requiring extensive personal behavioral history. 
Rather than relying solely on local data, new users can leverage  aggregated  behavioral patterns and risk discrimination capabilities while preserving privacy through the similarity-based transformation framework. 
This approach mitigates cold-start performance degradation by balancing security with usability and ensuring that even users with minimal session data can perform informed risk assessment.


\item \textit{Local Model Training and Federated Aggregation:}
For each user $\mathcal{U}_i$, the associated device maintains a local model $\mathcal{M}_i$, trained on per-session similarity vectors derived from that user’s sessions. 
The local training procedure, as outlined above,  is optimized to distinguish low-, medium-, and high-risk sessions, and begins once a sufficient number of labeled sessions are available to ensure stable similarity modeling and robust clustering. 
After local convergence, the model update $\Delta \mathcal{M}_i$, comprising the learned weight vector and bias term, is securely transmitted for federated aggregation. 
\acrshort{FLRBA2} uses a round-based synchronous update protocol, where eligible users participate in each global round, enabling consistent global model evolution while preserving user personalization.

\begin{figure}[!htpb]
 \centering
 \resizebox{\linewidth}{!}{
 \begin{tikzpicture}[scale=1.5, every node/.style={scale=1}]
 \node[minimum size=1.5cm] (server) at (0,0) {{
 \includegraphics[width=1.5cm]{server.png}
 }};
 \node [below=-2mm of server,xshift=1mm, align=center] {Central \\ Server};

 \foreach \i in {3,...,3}
 {
 \node[minimum size=1.5cm] (user\i) at (360/6 * \i:5) {{
 \includegraphics[width=0.8cm]{user.png}
 }};
 \node [align=center,below=-2mm of user\i,xshift=1mm] {User\\Device};
 \node[minimum size=1.5cm] (LocalModel\i) at (360/6 * \i:4) {{
 \includegraphics[width=0.8cm]{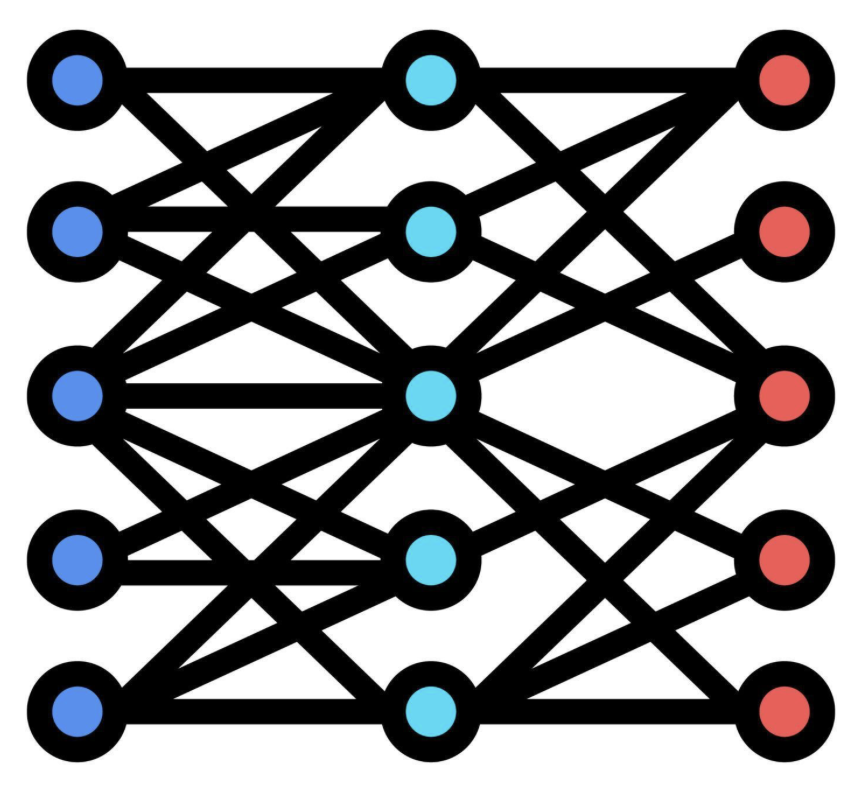}
 }};
 \node [align=center,below=-2mm of LocalModel\i,xshift=1mm] {Local\\Model};

 \draw[->,thick,blue] (server) -- (LocalModel\i) node[midway,above,sloped] {Global Model};

 \draw[->,thick,red,dashed] ($(LocalModel\i.east)+(0,-0.2)$) -- ($(server.west)+(0,-0.2)$) node[midway,below,sloped] {Noisy Model Update};
 }
 \end{tikzpicture}}
 \caption{Federated Learning with \gls{DP} in \acrshort{FLRBA2}}
\end{figure}

\item \textit{Differential Privacy in Model Updates:} 
Privacy preservation in \gls{FL} requires careful protection of individual user data while maintaining model utility. Each device $\mathcal{U}_i$ applies \gls{DP} to its model update $\Delta \mathcal{M}_i$ before transmission to the central server. This privacy-preserving transformation adds calibrated Gaussian noise based on a predefined privacy budget $\epsilon$:
\begin{align*}
  \Delta \mathcal{M}_i' = \Delta \mathcal{M}_i + \mathcal{N}(0, \sigma^2),  
\end{align*}
where $\mathcal{N}(0, \sigma^2)$ is Gaussian noise with variance $\sigma^2$. The noise variance is determined using the moments accountant method, providing tight privacy loss bounds compared to traditional composition. This noise calibration ensures individual user data remains computationally indistinguishable in the aggregated model while preserving sufficient utility for effective learning. The \gls{DP} guarantee provides formal protection against honest-but-curious and malicious adversaries, meeting stringent privacy requirements for authentication systems.
\end{enumerate}
\subsection{Federated Learning Aggregation and Global Model Update}
The central server orchestrates the \gls{FL} process, aggregating differentially private model updates from all participating devices. This transforms individual device insights into collective intelligence while preserving user privacy.
The server aggregates the differentially private model updates $\Delta \mathcal{M}_i'$ to compute an overall model improvement:
\begin{align*}
\Delta \mathcal{M} = \frac{1}{n} \left(\sum_{i=1}^{n} \Delta \mathcal{M}_i'\right)= \frac{1}{n} \left(\sum_{i=1}^{n} \Delta \mathcal{M}_i \right)+ \mathcal{N}(0, \frac{\sigma^2}{n}).
\end{align*}
The global model is then updated using the aggregated improvement:
\begin{align*}
\mathcal{M}_{t+1} = \mathcal{M}_t + \Delta \mathcal{M}
\end{align*}
This refined global model $\mathcal{M}_{t+1}$ is redistributed to all participating devices, enhancing their local risk assessment capabilities. Through this federated aggregation, devices benefit from collective insights while maintaining data locality. This approach particularly benefits new devices by providing well-informed initial thresholds and parameters to effectively address cold-start limitations.
\begin{enumerate}[topsep=1ex, itemsep=1ex, wide, font=\itshape, labelwidth=!, labelindent=0pt, label*=D.\arabic*.]

\item \textit{Privacy–Utility Trade-Off and Consistency in Model Training:}
\acrshort{FLRBA2} balances privacy–utility trade-offs through systematic $(\epsilon, \delta)$-differential privacy calibration, where $\epsilon$ bounds cumulative privacy loss and $\delta$ denotes the failure probability. 
Each device applies the Gaussian mechanism with bounded $\ell_2$-sensitivity, and the noise variance $\sigma^2$ is calibrated using the moments accountant method~\cite{9082603}. 
The moments accountant tracks log-moments of the privacy loss random variable across training rounds, providing tighter cumulative privacy bounds than classical composition. 
This enables \gls{FLRBA2} to inject the minimum noise necessary to satisfy target privacy budgets, thereby preserving accuracy while guaranteeing formal privacy protection. 
This principled approach ensures consistent global convergence under regulatory privacy constraints. Section~\ref{sec:performance_evaluation} reports empirical noise ranges explored per modality to characterize privacy–utility trade-offs in practice.

\end{enumerate}

\section{Security Analysis}
\label{sec:security_analysis}

This section provides a comprehensive security analysis of \acrshort{FLRBA2} within the Random Oracle Model, establishing formal guarantees for adaptive authentication, privacy, and correctness. Our analysis employs game-based proofs to demonstrate security under standard cryptographic assumptions.

\subsection{Security Assumptions}

The security of \acrshort{FLRBA2} is based on the following well-established assumptions:

\begin{itemize}
\item \textbf{Random Oracle Model (ROM):} We model the MAC function as a random oracle $\mathcal{H} : \{0,1\}^* \to \{0,1\}^\lambda$ that produces pseudorandom outputs, ensuring unpredictability for adversaries without access to the MAC key.

\item \textbf{Existentially Unforgeable MACs:} The message authentication code satisfies existential unforgeability under chosen-message attacks (EUF-CMA), preventing efficient adversaries from forging valid authentication tags.

\item \textbf{Differential Privacy Guarantee:} The Gaussian mechanism provides $(\epsilon, \delta)$-\gls{DP} by ensuring statistical indistinguishability between outputs from neighboring datasets.
\item \textbf{Bounded Sensitivity:} The $\ell_2$-sensitivity of similarity vector computations is bounded by constant $\Delta_2 \leq C$, enabling precise calibration of \gls{DP} noise parameters. This assumption is satisfied by our similarity transformation framework, where each similarity score $\text{Sim}_{F_i} \in [0,1]$ (Section~\ref{sec:similarity_computation}), yielding $\ell_2$-sensitivity $\Delta_2 \leq \sqrt{n}$ for $n$-dimensional similarity vectors.
\end{itemize}

\subsection{Adaptive Authentication Security}

We first establish that \acrshort{FLRBA2} prevents impersonation attacks through adaptive authentication mechanisms.

\begin{game}[\textbf{Adaptive Authentication Security}]
Let \(\Pi\) be an authentication protocol that uses a secret MAC key \(k\) and a random oracle \(\mathcal{H} : \{0,1\}^* \to \{0,1\}^\lambda\). The protocol is said to be \((\epsilon, q, \delta)\)-secure if,  any \gls{PPT} adversary \(\mathcal{A}\) making at most \(q\) queries,  has at most  advantage  \(\epsilon\) in the following adaptive authentication security game:
\begin{itemize}
    \item  \textbf{Setup.} The challenger samples a target user profile \(\mathcal{H}^* = \{\bm{s}_1, \dots, \bm{s}_m\} \leftarrow \mathcal{D}_{\text{profile}}\), and computes the reference vector (centroid) as \(\text{Centroid}^* = \frac{1}{m} \sum_j \bm{s}_j\).

    \item  \textbf{Challenge Phase.} \(\mathcal{A}\) may adaptively issue the following types of queries. A query \texttt{Observe}(\(j\)) returns the authenticated session tuple \((\bm{s}_j, \text{ctx}_j, 1)\) for the \(j\)-th session of the target user \(\mathcal{U}^*\). A query \texttt{Test}(\(\bm{Sim}, \mathcal{U}_i, \text{ctx}\)), with \(\mathcal{U}_i \ne \mathcal{U}^*\), returns the predicate
\(\mathbb{1}\left[\| \bm{Sim} - \text{Centroid}_i \|_2 \leq \mathcal{H}(\text{MAC}_k(\text{Centroid}_i) \| \text{ctx} \| i)\right]\),
where \(\text{Centroid}_i = \frac{1}{m} \sum_{j=1}^m \bm{s}_{i,j}\) is the centroid of enrolled vectors for user \(\mathcal{U}_i\). A query \texttt{RO}(\(x\)) returns \(\mathcal{H}(x)\); if \(x\) has not been queried before, a uniformly random value from \(\{0,1\}^\lambda\) is sampled and fixed for future queries.

    \item  \textbf{Guess.} \(\mathcal{A}\) outputs a candidate pair \((\bm{Sim}^*, \text{ctx}^*)\) such that \(\|\bm{Sim}^* - \bm{s}_j\|_2 > \delta\) for all \(j\). It wins if
\(\| \bm{Sim}^* - \text{Centroid}^* \|_2 \leq \mathcal{H}(\text{MAC}_k(\text{Centroid}^*) \| \text{ctx}^* \| i^*)\),
where \(i^*\) is the index of the target user \(\mathcal{U}^*\).
\end{itemize}
The advantage of \(\mathcal{A}\) in the above game is defined as:
\[\text{Adv}^{\text{adapt-auth}}_{\Pi,\mathcal{A}}(\lambda) := \left|\Pr[\text{Game}_{\mathcal{A}}^{\text{adapt-auth}}(\lambda) = 1] - 2^{-\lambda}\right|.\]
\end{game}

\begin{thm}[\textbf{Adaptive Authentication Security in  \acrshort{FLRBA2}}]
Let \(\Pi\) be the \acrshort{FLRBA2} protocol using a \((\epsilon_{\mathrm{MAC}}, q)\)-unforgeable MAC and a random oracle \(\mathcal{H} \colon \{0,1\}^* \to \{0,1\}^\lambda\). Then \(\Pi\) is \((\epsilon, q, \delta)\)-secure against adaptive authentication, where 
\[
\epsilon \leq \epsilon_{\mathrm{MAC}} + \frac{q_H + 1}{2^\lambda}
\]
\end{thm}

\begin{proof}
Suppose a PPT adversary \(\mathcal{A}\) succeeds in the adaptive authentication game with advantage \(\epsilon\). We construct a reduction adversary \(\mathcal{B}\) that breaks either MAC unforgeability or the random oracle assumption. \(\mathcal{B}\) is given access to a MAC oracle \(\text{MAC}_k(\cdot)\) and simulates the game for \(\mathcal{A}\) by sampling a target profile \(\mathcal{H}^* = \{s_1, \ldots, s_m\}\), computing the reference vector \(\text{Centroid}^* = \frac{1}{m} \sum_j s_j\), and initializing a lazy-sampled table \(H\) to simulate the random oracle \(\mathcal{H}\). For each \texttt{Test} query on user \(\mathcal{U}_i \ne \mathcal{U}^*\), \(\mathcal{B}\) computes \(\text{Centroid}_i\), queries \(\tau_i = \text{MAC}_k(\text{Centroid}_i)\), and returns the predicate \(\mathbb{1}[\|\bm{Sim} - \text{Centroid}_i\|_2 \leq H[\tau_i \| \text{ctx} \| i]]\). Random oracle queries are answered uniformly at random and cached in \(H\). Upon challenge \((\bm{Sim}^*, \text{ctx}^*)\), \(\mathcal{B}\) computes \(\tau^* = \text{MAC}_k(\text{Centroid}^*)\). If \(\tau^*\) was not previously queried, \(\mathcal{B}\) outputs \((\text{Centroid}^*, \tau^*)\) as a MAC forgery. Otherwise, \(\mathcal{A}\) must have guessed \(\mathcal{H}(\tau^* \| \text{ctx}^* \| i^*)\) without querying the oracle, which occurs with probability at most \(\frac{1}{2^\lambda}\) or \(\frac{q_H}{2^\lambda}\) after \(q_H\) queries. Therefore, the total success probability is bounded by \(\epsilon \leq \epsilon_{\mathrm{MAC}} + \frac{q_H + 1}{2^\lambda}\).
\end{proof}
\subsection{Privacy Analysis}

We next demonstrate that \acrshort{FLRBA2} provides strong privacy guarantees through \gls{DP} mechanisms.

\begin{game}[\textbf{Federated Differential Privacy}]\label{game:fed-dp}
Let \(\Pi\) be a \gls{FL} protocol employing a Gaussian mechanism for privacy and a message authentication code (MAC) for integrity. The protocol satisfies \((\epsilon, \delta, T)\)-federated \gls{DP} if no \gls{PPT} adversary \(\mathcal{A}\), participating in at most \(T\) training rounds, can distinguish between two protocol executions that differ only in the local data of a single target user \(\mathcal{U}^*\). The game proceeds as follows:

\begin{enumerate}
  \item \textbf{Setup.} The challenger \(\mathcal{C}\) samples \(n\) users \(\{\mathcal{U}_1, \dots, \mathcal{U}_n\}\), selects a target user \(\mathcal{U}^*\), and constructs two adjacent datasets \(D_0\) and \(D_1\) differing only in \(\mathcal{U}^*\)'s behavioral data. A secret MAC key \(k\) and a random oracle \(\mathcal{H} : \{0,1\}^* \to \{0,1\}^\lambda\) are sampled.

  \item \textbf{Corruption Phase.} \(\mathcal{A}\) adaptively corrupts up to \(c < n/2\) users (excluding \(\mathcal{U}^*\)), gaining access to their local data and computations.

\item \textbf{Challenge.} The challenger \(\mathcal{C}\) samples a hidden bit \(b \in \{0,1\}\) and executes the protocol \(\Pi\) on dataset \(D_b\) for \(T\) \gls{FL} rounds. During each round, every user perturbs their local model update using Gaussian noise \(\mathcal{N}(0, \sigma^2 I)\) and transmits a MAC-authenticated version of the update to the central server. The adversary observes the global model after each round, the noisy updates and corresponding MAC tags of the corrupted users, and the overall protocol transcript, excluding the plaintext updates originating from the target user \(\mathcal{U}^*\).
  \item \textbf{Guess.} \(\mathcal{A}\) outputs a bit \(b'\), attempting to determine whether the execution used \(D_0\) or \(D_1\).
\end{enumerate}

The adversary’s advantage is defined as:
\[
\text{Adv}^{\text{fed-dp}}_{\Pi,\mathcal{A}}(\lambda) := \left| \Pr[b' = b] - \frac{1}{2} \right|.
\]
\end{game}

\begin{thm}[\textbf{Federated Differential Privacy in  \acrshort{FLRBA2}}]
Let \(\Pi\) be the FL-RBA\(^2\) protocol incorporating:
(i) an \((\epsilon_{\mathrm{MAC}}, q)\)-unforgeable MAC,
(ii) a random oracle \(\mathcal{H} : \{0,1\}^* \to \{0,1\}^\lambda\), and
(iii) a Gaussian mechanism \(\mathcal{N}(0, \sigma^2)\) with \(\sigma = \frac{2\Delta_2 \sqrt{2T \ln(2/\delta)}}{\epsilon}\), where \(\Delta_2 \leq C\) bounds the sensitivity. Then \(\Pi\) satisfies \((\epsilon + \epsilon_{\mathrm{MAC}}, \delta, T)\)-federated \gls{DP} with
\[
\text{Adv}^{\text{fed-dp}}_{\Pi,\mathcal{A}}(\lambda) \leq \epsilon + \epsilon_{\mathrm{MAC}} + \delta + \text{negl}(\lambda).
\]
\end{thm}

\begin{proof}
We bound the advantage of adversary \(\mathcal{A}\) through a sequence of hybrid games. In \textit{Game 0}, the protocol is executed on dataset \(D_b\) using real MAC tags and Gaussian noise. In \textit{Game 1}, all MAC values are replaced with simulated outputs from a random oracle \(\mathcal{H}\); if \(\mathcal{A}\) distinguishes this change with probability greater than \(\epsilon_{\mathrm{MAC}}\), then a reduction can break MAC unforgeability. In \textit{Game 2}, the Gaussian mechanism is substituted with an ideal \((\epsilon, \delta)\)-differentially private mechanism, altering \(\mathcal{A}\)'s view by at most \(\delta\). In \textit{Game 3}, this ideal mechanism guarantees that the transcript is \(\epsilon\)-indistinguishable across \(T\) rounds, given per-user sensitivity bounded by \(\Delta_2 \leq C\) and noise scale \(\sigma = \frac{2C \sqrt{2T \ln(2/\delta)}}{\epsilon}\). By the triangle inequality over these transitions, \(\mathcal{A}\)'s total distinguishing advantage is at most \(\epsilon + \epsilon_{\mathrm{MAC}} + \delta + \text{negl}(\lambda)\).
\end{proof}

\subsection{Correctness Analysis}

Finally, we prove that \acrshort{FLRBA2} ensures the integrity of federated model aggregation.

\begin{game}[\textbf{Correctness}]\label{game:correctness}
Let $\Pi$ be a \gls{FL} protocol using MAC-based authentication. We define the game $\mathsf{Game}_{\text{Correct}}^{\mathcal{A},\Pi}(\lambda)$ between a challenger $\mathcal{C}$ and an adversary $\mathcal{A}$ as follows:

\begin{enumerate}
  \item \textbf{Setup.} $\mathcal{C}$ generates secret keys $\{k_i\}_{i=1}^n$ for $n$ users and computes authenticated local updates $\{(\Delta\mathcal{M}_i, \tau_i)\}_{i=1}^n$ where $\tau_i \leftarrow \text{MAC}_{k_i}(\Delta\mathcal{M}_i \| t_i \| i)$. The correct global model is defined as $\mathcal{M} := \frac{1}{n} \sum_{i=1}^n \Delta\mathcal{M}_i$.

  \item \textbf{Adversarial Tampering.} $\mathcal{A}$ outputs a modified set of updates $\{(\Delta\mathcal{M}_i^*, \tau_i^*, t_i^*, i^*)\}$, resulting in $\mathcal{M}^* := \frac{1}{|S|} \sum_{i \in S} \Delta\mathcal{M}_i^*$ where $S$ includes all updates that pass verification.

  \item \textbf{Verification.} $\mathcal{C}$ accepts each update if $\mathsf{Verify}_{k_{i^*}}(\Delta\mathcal{M}_i^* \| t_i^* \| i^*, \tau_i^*) = 1$ and the timestamp is fresh.
\end{enumerate}

The protocol $\Pi$ satisfies aggregation correctness if:
\[
\Pr\left[ \mathcal{M}^* \neq \mathcal{M} \wedge \forall i,~ \mathsf{Verify}_{k_i}(\Delta\mathcal{M}_i^* \| t_i^* \| i, \tau_i^*) = 1 \right] \leq \text{negl}(\lambda)
\]
for all \gls{PPT} adversaries $\mathcal{A}$.
\end{game}

\begin{thm}[\textbf{Correctness of FL-RBA}$^2$]
Let $\Pi$ be the \acrshort{FLRBA2} protocol using a secure message authentication code $\mathcal{MAC} = (\text{Gen}, \text{Tag}, \text{Verify})$. Then for any PPT adversary $\mathcal{A}$ against correctness, there exists a PPT adversary $\mathcal{B}$ against MAC unforgeability such that:
\[
\text{Adv}_{\Pi,\mathcal{A}}^{\text{correct}}(\lambda) \leq n \cdot \text{Adv}_{\mathcal{MAC},\mathcal{B}}^{\text{mac}}(\lambda) + \text{negl}(\lambda)
\]
\end{thm}

\begin{proof}
We construct an adversary $\mathcal{B}$ that breaks MAC unforgeability by simulating the correctness game with a challenge key $k^*$. $\mathcal{B}$ randomly selects a user index $j^* \in \{1, \dots, n\}$ to embed the unforgeability challenge. For all $i \neq j^*$, it generates keys $k_i \leftarrow \mathsf{Gen}(1^\lambda)$ and simulates local updates $\{\Delta\mathcal{M}_i\}$ and their MAC tags.
To simulate the authentication phase, $\mathcal{B}$ queries the MAC oracle for $(\Delta\mathcal{M}_{j^*} \| t_{j^*} \| j^*)$ and receives $\tau_{j^*}$. It sends all authenticated updates to $\mathcal{A}$.
If $\mathcal{A}$ outputs a forged update $(\Delta\mathcal{M}_{j^*}^*, \tau_{j^*}^*, t_{j^*}^*, j^*)$ such that $\mathsf{Verify}_{k^*}(\Delta\mathcal{M}_{j^*}^* \| t_{j^*}^* \| j^*, \tau_{j^*}^*) = 1$ and either the message or timestamp differs, then $\mathcal{B}$ outputs $(m^*, \tau^*)$ as a valid forgery. Otherwise, $\mathcal{B}$ aborts.
Since $\mathcal{A}$ causes a deviation $\mathcal{M}^* \neq \mathcal{M}$ only by modifying at least one update with a forged MAC, and $\mathcal{B}$ guesses the correct target user with probability $1/n$, the reduction holds with:
\[
\text{Adv}_{\Pi,\mathcal{A}}^{\text{correct}}(\lambda) \leq n \cdot \text{Adv}_{\mathcal{MAC},\mathcal{B}}^{\text{mac}}(\lambda) + \text{negl}(\lambda)
\]

Thus, correctness of \acrshort{FLRBA2} reduces to MAC unforgeability.
\end{proof}

\subsection{Security Guarantees Summary}

Our analysis establishes three key security properties under standard cryptographic assumptions: robust authentication through adaptive threshold-based mechanisms and MAC protection preventing impersonation attacks with negligible advantage $\mathrm{negl}(\lambda)$ (ROM, EUF-CMA MAC); strong privacy via \gls{DP} mechanisms ensuring statistically indistinguishable user contributions in \gls{FL}, achieving $(\epsilon + \mathrm{negl}(\lambda), \delta)$-DP guarantees through bounded sensitivity ($\Delta_2 \leq \sqrt{n}$) of our similarity transformation framework; and aggregation integrity through MAC-based authentication preventing adversarial tampering with negligible probability (EUF-CMA MAC). 
These properties make \acrshort{FLRBA2} resilient against impersonation, inference, model inversion, and tampering attacks, establishing suitability for privacy-critical adaptive authentication and \gls{FL} applications.

\section{Implementation and Performance Evaluation}
\label{sec:performance_evaluation}

This section presents a comprehensive evaluation of \acrshort{FLRBA2}, demonstrating its effectiveness in privacy-preserving federated risk classification across heterogeneous authentication modalities. We evaluate our framework on three distinct datasets representing different categories of authentication features: keystroke dynamics, mouse behavioral patterns, and contextual login data. Table~\ref{tab:flrba2_datasets} provides a detailed overview of the components used across all modalities.
\begin{table*}[htbp]

\renewcommand{\arraystretch}{1.2}
\scriptsize
\caption{Overview of \acrshort{FLRBA2} Components Across Keyboard, Mouse, and Contextual Modalities}
\label{tab:flrba2_datasets}
\resizebox{\textwidth}{!}{
\begin{tabular}{|p{0.09\linewidth}|p{0.1\linewidth}|p{0.26\linewidth}|p{0.20\linewidth}|p{0.18\linewidth}|p{0.17\linewidth}|}
\hline
\textbf{Modality} & \textbf{Dataset} & \textbf{Features Used} & \textbf{Similarity Modeling} & \textbf{Risk Labeling Method} & \textbf{Training \& Aggregation} \\
\hline

\textbf{Keystroke} & 
CMU Keystroke Dynamics Benchmark~\cite{cmu_keystroke} &
31 timing features per password: 
\begin{minipage}[t]{\linewidth} \begin{MyItemize}
  \item 11 Hold times (key press–release)
  \item 10 Down–Down latencies (between key presses)
  \item 10 Up–Down latencies (key release to next press)
\end{MyItemize} \end{minipage} &
\begin{minipage}[t]{\linewidth} \begin{MyItemize}
\item FastDTW to user-specific DBA centroid
\item One similarity score per test session
\item Normalized via per-user MaxDTW

\end{MyItemize} \end{minipage} &
\begin{minipage}[t]{\linewidth} \begin{MyItemize}
  \item KMeans clustering ($k=3$) on mean similarity
  \item Centroids sorted for Low–Med–High relabeling
\end{MyItemize} \end{minipage} &
\begin{minipage}[t]{\linewidth} \begin{MyItemize}
  \item Logistic Regression (L2, class-balanced)
  \item 3-round FedAvg across users
  \item Optional Gaussian DP noise ($\sigma \in [0, 1]$)
\end{MyItemize} \end{minipage} \\
\hline

\textbf{Mouse} & 
Balabit Mouse Dynamics Challenge~\cite{balabit_mouse} &
Session sequences from raw events:
\begin{minipage}[t]{\linewidth} \begin{MyItemize}
  \item Spatial: $(x, y)$ trajectories
  \item Temporal: velocity, acceleration
  \item State: button, interaction type
\end{MyItemize} \end{minipage} &
\begin{minipage}[t]{\linewidth} \begin{MyItemize}
  \item FastDTW on sequence-level coordinate series
  \item Top-5 reference matches
  \item Normalized via per-user MaxDTW
\end{MyItemize} \end{minipage} &
\begin{minipage}[t]{\linewidth} \begin{MyItemize}
  \item KMeans ($k=3$) on top-$k$ mean similarity\textsuperscript{*}
  \item Relabeled by centroid similarity ranking
\end{MyItemize} \end{minipage} &
\begin{minipage}[t]{\linewidth} \begin{MyItemize}
  \item Logistic Regression per user
  \item FedAvg aggregation over 3 rounds
  \item Optional Gaussian DP ($\sigma \in [0, 1]$)
\end{MyItemize} \end{minipage} \\
\hline

\textbf{Contextual} & 
RBA Dataset (Kaggle)~\cite{rba_dataset} (Augmented with GeoLite2 data~\cite{geolite2city}) &
24 engineered features from each login:
\begin{minipage}[t]{\linewidth} \begin{MyItemize}
  \item Geolocation: city, region, country, lat/lon
  \item Device and software: OS/browser name/version, device type
  \item Network: IP, ASN, ASN category, IsBenign
  \item Temporal: login time, hour, weekday, 7-day frequency
  \item Dynamics: version drift, login gap, agent string components
\end{MyItemize} \end{minipage} &
\begin{minipage}[t]{\linewidth} \begin{MyItemize}
  \item Per-feature similarity vectors (categorical, geospatial, binary)
  \item Haversine distance on coordinates
  \item Min-Max scaling across features
\end{MyItemize} \end{minipage} &
\begin{minipage}[t]{\linewidth} \begin{MyItemize}
  \item KMeans clustering ($k=2$--$3$)
  \item Relabeling via centroid norm (low/med/high)
\end{MyItemize} \end{minipage} &
\begin{minipage}[t]{\linewidth} \begin{MyItemize}
  \item Logistic Regression (L2, maxiter=200)
  \item FedAvg across users (3 rounds)
  \item Optional DP via Gaussian noise ($\sigma \in [0, 1]$)
\end{MyItemize} \end{minipage} \\
\hline
\end{tabular}}
$\textsuperscript{*}$The first $k{=}3$ refers to the number of clusters used in KMeans; the second $k{=}5$ refers to the top-$k$ DTW similarities used to construct each similarity vector.
\vspace{-0.2cm}
\end{table*}

\subsection{Experimental Setup and Datasets}
\label{subsec:datasets}

We conduct a comprehensive evaluation of \acrshort{FLRBA2} across three authentication modalities: keystroke dynamics, mouse dynamics, and contextual features. We leverage established benchmark datasets that span diverse behavioral dimensions, user populations, and privacy-utility trade-offs essential for real-world federated authentication.
For keystroke-based behavioral modeling, we employ the \textit{CMU Keystroke Dynamics Benchmark Dataset}~\cite{cmu_keystroke}, a widely recognized benchmark for fine-grained temporal biometrics. The dataset comprises data from 51 users, each instructed to type the strong, fixed password ``\texttt{.tie5Roanl}'' 400 times across 8 sessions. Each session includes structured metadata fields such as \texttt{subject}, \texttt{sessionIndex}, and \texttt{rep}, along with 31 timing features extracted from the fourth column onward. These features consist of 11 \textit{Hold Time (H)} values representing individual key press durations, 10 \textit{Down-Down Latency (DD)} values capturing inter-key press intervals, and 10 \textit{Up-Down Latency (UD)} values measuring the delay between key release and subsequent press. This temporal feature set enables highly discriminative user profiling while supporting privacy-preserving federated learning through personalized similarity modeling.

To evaluate mouse dynamics, we use the \textit{Balabit Mouse Dynamics Challenge Dataset}~\cite{balabit_mouse}, which records detailed mouse interaction data from 10 users, including event timestamps, cursor positions, button states, and event types. The raw sessions are preprocessed to remove invalid entries (e.g., missing or zero-valued coordinates) and then merged into a unified format containing fields such as \texttt{user\_id}, \texttt{session\_id}, \texttt{x}, \texttt{y}, \texttt{button}, and \texttt{state}. Each user's sessions are grouped to extract raw $(x, y)$ cursor trajectories and compute high-order behavioral indicators including velocity and acceleration patterns. This feature space captures spatial-kinetic behavior with sufficient variability for personalized similarity modeling.

For contextual authentication, we adopt the large-scale \textit{Risk-Based Authentication Dataset}~\cite{rba_dataset}, which contains over 3.5 million login events spanning 97,854 users. The dataset is enriched with IP-based geolocation metadata derived from the GeoLite2 City Database~\cite{geolite2city}, and processed through a multi-stage contextual modeling pipeline. User agent strings are parsed into granular subcomponents such as \texttt{UA\_OS}, \texttt{UA\_Arch}, and \texttt{UA\_Engine}, and browser/OS version fields are extracted using regular expressions from both the user agent and dedicated metadata fields. IP addresses are resolved to geographic coordinates and Autonomous System Numbers (ASNs) using the \texttt{geoip2} library, while additional features such as \texttt{Is Benign IP}, round-trip time, and temporal login patterns (e.g., time-of-day, frequency) are incorporated to enhance contextual granularity. Users are excluded if they exhibit fewer than two login events, fewer than three distinct contextual profiles, or low behavioral variance. After preprocessing and normalization, each login event is represented by a 24-dimensional contextual vector spanning geolocation, device fingerprinting, network behavior, temporal context, and behavioral dynamics. These vectors form the basis for similarity computation and federated risk modeling in the \acrshort{FLRBA2} framework.
\subsection{Similarity Evaluation and Local Training}
The pipeline first groups sessions per user and partitions them into reference and test sets. For keystroke data, a user-specific reference centroid is computed via DTW barycenter averaging over reference sessions to capture typical behavior. 
For mouse data, the high variability of mouse trajectories makes centroid averaging unstable. 
Instead, each test session is compared to all reference sessions using FastDTW, 
and the top-$k$ most similar matches form a $k$-dimensional similarity vector $\text{Sim}_F$. 
We empirically set $k=5$, which provides the best balance between robustness and efficiency: 
lower values were too noise-sensitive, while higher values added redundant information 
without improving classification performance. 
This choice ensures the similarity vector remains tractable for local model training.
For contextual features, similarity vectors are computed per user across consecutive logins. Categorical similarity is based on exact matches (e.g., hour of day, day of week, device type, browser/OS name and version, ASN, user agent components, geolocation tags), geographic similarity uses haversine distance between coordinates, and temporal similarity is derived from a normalized 7-day login frequency. Each feature contributes a similarity score in $[0,1]$.
 Similarity vectors undergo Min-Max scaling; users with low variance or insufficient data are excluded. Risk levels are inferred via KMeans clustering on similarity vectors ($k{=}3$ by default; $k{=}2$ optional for stable users), with cluster labels ordered by centroid magnitude for consistent risk stratification.
Each user trains a local L2-regularized logistic regression model (200 iterations, class balancing) on similarity vectors labeled by KMeans. Federated Averaging (FedAvg) is performed over 3 rounds to produce a global model, aggregating locally trained coefficients and intercepts with zero-padding for missing dimensions. The choice of 3 rounds reflects a practical balance: logistic regression converges quickly, and empirical evaluation showed that performance stabilized within three rounds, while additional rounds provided negligible gains but higher communication overhead. \gls{DP} is supported by injecting Gaussian noise  into model parameters \(
\theta_i^{\text{noisy}} = \theta_i + \mathcal{N}(0, \sigma^2)
\).
Final evaluation uses macro-averaged metrics (accuracy, precision, recall, F1-score). The \acrshort{FLRBA2} pipeline unifies engineered features, unsupervised risk labeling, supervised training, and federated aggregation to enable privacy-preserving, adaptive, and scalable risk-based authentication. Table~\ref{tab:flrba2_datasets} summarizes learning components by modality.

\subsection{Experimental Results}
\label{subsec:results}

\subsubsection{Baseline Performance Analysis}
To evaluate the core effectiveness of \acrshort{FLRBA2} prior to integrating privacy-preserving mechanisms, we conducted a baseline assessment across keystroke, contextual, and mouse modalities with \gls{DP} disabled. This configuration isolates the system's ability to detect high-risk behavior based solely on similarity-based modeling and federated training. The evaluation employed a train/test split ratio of 0.7, allocating 70\% of the data for model training and 30\% for performance assessment.

Among all modalities, contextual features demonstrated the highest accuracy (0.9698) and strong sensitivity to malicious activity, achieving an F1-score of 0.8031 and a recall of 0.9606 for the high-risk class.  This highlights the modality’s robustness in identifying adversarial access patterns---such as anomalous IP addresses, device fingerprints, and environmental inconsistencies. The discrete and relatively stable nature of contextual attributes supports effective separation between benign and suspicious sessions.
Mouse dynamics, despite inherent variability and sensor noise, achieved a high-risk F1-score of 0.7907 with a precision of {0.8947}. These results indicate that although accuracy may be lower due to behavioral fluctuations, this modality remains highly effective at reducing false positives in malicious session detection.
The keystroke modality yielded an F1-score of 0.7589 for high-risk detection, with a recall of 0.7944 and  high  accuracy of 0.7943. This reflects the discriminative power of temporal typing patterns, which are difficult to replicate and serve as stable behavioral signatures under adversarial conditions.
\begin{figure}[!h]

\hfill
\begin{tikzpicture}[scale=0.93]

\begin{axis}[
    width=\linewidth,
    height=6.5cm,
    ybar,
    bar width=10pt,
    enlarge x limits=0.2,
    ymin=0.0, ymax=1,
    ylabel={Performance Metrics},
     yticklabel style={font=\scriptsize},
    ylabel style={yshift=0em},
    xtick=data,
    symbolic x coords={Keystroke, Mouse, Contextual},
    xticklabel style={font=\small},
    nodes near coords,
    every node near coord/.append style={font=\tiny, yshift=2pt},
    legend style={draw=none},
    axis x line*=bottom,
    axis y line*=left,
    tick align=inside,
    tick style={thin},
]

\addplot+[ybar, fill=blue] coordinates {(Keystroke, 0.7943) (Mouse, 0.7083) (Contextual, 0.9698)};
\addplot+[ybar, fill=orange] coordinates {(Keystroke, 0.7265) (Mouse, 0.8947) (Contextual, 0.7323)};
\addplot+[ybar, fill=gray] coordinates {(Keystroke, 0.7944) (Mouse, 0.7083) (Contextual, 0.9609)};
\addplot+[ybar, fill=yellow] coordinates {(Keystroke, 0.7589) (Mouse, 0.7907) (Contextual, 0.8031)};

\end{axis}

\end{tikzpicture}

\vspace{-1.4em}
\hfill
\scriptsize
\renewcommand{\arraystretch}{1.2}
\resizebox{0.988\linewidth}{!}{\begin{tabular}{l
                |>{\centering\arraybackslash}m{1.5cm}
                |>{\centering\arraybackslash}m{1.5cm}
                |>{\centering\arraybackslash}m{1.5cm}|}
&&&\\
\hline
\multicolumn{1}{|l|}{\legendcolorbox{blue} Accuracy} & {0.7943} & 0.7083 & \textbf{0.9698} \\ \hline
\multicolumn{1}{|l|}{\legendcolorbox{orange} Precision} & 0.7265 & \textbf{0.8947} & 0.7323 \\ \hline
\multicolumn{1}{|l|}{\legendcolorbox{gray} Recall} & 0.7944 & 0.7083 & \textbf{0.9606} \\ \hline
\multicolumn{1}{|l|}{\legendcolorbox{yellow} F1-score }& 0.7589 & 0.\textbf{7907} & \textbf{0.8031} \\
\hline
\end{tabular}}

\caption{Baseline Modality-Wise Performance of \acrshort{FLRBA2} for High-Risk User Detection}
\label{fig:class2_barlegend}\vspace{-0.5em}
\end{figure}
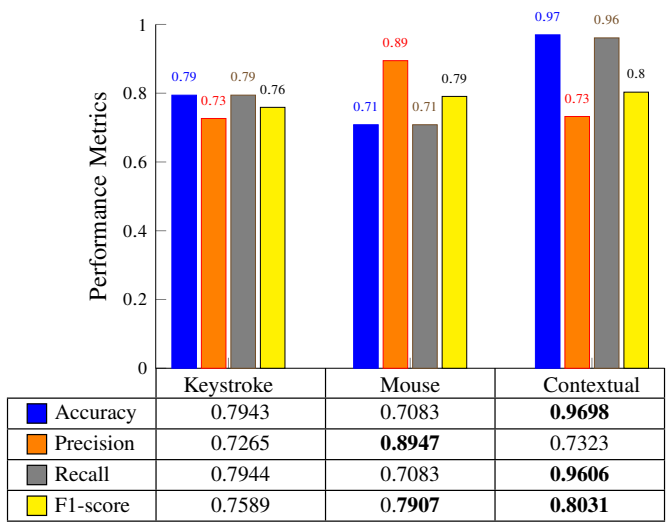

Overall, these results, summarized in Figure \ref{fig:class2_barlegend}, confirm that \acrshort{FLRBA2} achieves its primary security objective: accurate and modality-agnostic identification of high-risk users, through unsupervised similarity modeling, cluster-driven risk labeling, and federated logistic regression, even in the absence of \gls{DP} noise injection.

\subsubsection{Differential Privacy Impact Analysis}
Our modality-specific DP analysis reveals distinct privacy-utility trade-offs, as detailed in Table~\ref{tab:dp-modality-integrated}. The privacy budget calculation $\varepsilon = \frac{\Delta \sqrt{2 \ln(1.25/\delta)}}{\sigma} \approx \frac{4.845}{\sigma}$ (with $\Delta = 1.0$, $\delta = 10^{-5}$) remains mathematically consistent across modalities, but practical noise parameter selection requires careful calibration due to differential sensitivity to additive Gaussian perturbations.

 \begin{table*}[ht]
\centering
\scriptsize
\caption{Modality-Specific Differential Privacy Configuration, Robustness, and Dataset Characteristics}
\label{tab:dp-modality-integrated}
\resizebox{\textwidth}{!}{
\begin{tabular}{|p{0.07\linewidth}|p{0.12\linewidth}|p{0.14\linewidth}|p{0.17\linewidth}|p{0.15\linewidth}|p{0.34\linewidth}|}
\hline
\textbf{Modality} & \textbf{Noise Sensitivity} & \textbf{Optimal Noise Scale ($\boldsymbol{\sigma}$)} & \textbf{Observed Privacy Budget ($\boldsymbol{\varepsilon}$)} & \textbf{Achievable Privacy Level} & \textbf{Robustness Justification} \\
\hline
\textbf{Keystroke} & \textbf{High}: utility degrades sharply for $\sigma \geq 0.5$ & $[0.01,\ 1.0]$ & $[4.8,\ 484]$ & $\varepsilon \leq 4.8$ & Low-variance temporal features (e.g., hold/latency intervals) combined with a small user pool ($N=51$) increase overfitting risk and reduce tolerance to noise injection.\\
\hline
\textbf{Mouse} & \textbf{Medium}: tolerable degradation up to $\sigma \approx 2.0$ & $[0.02,\ 2.0]$ & $[2.4,\ 242]$ & $\varepsilon \leq 2.4$ & Moderate robustness due to spatial-kinetic variability (e.g., velocity, acceleration). Although limited user count ($N=10$) reduces generalization, the high signal variance increases resilience to DP noise. \\
\hline
\textbf{Contextual} & \textbf{Low}: negligible utility loss for $\sigma \leq 2.0$ & $[0.5,\ 8.0]$ & $[0.61,\ 9.7]$ & $\varepsilon \leq 1$ & High-entropy, high-cardinality features (e.g., device ID, OS, timezone) exhibit weak identity coupling. The large user base ($N=97,\!854$) enables strong generalization and noise robustness under DP. \\
\hline
\end{tabular}
}
\vspace{0.5cm}
\end{table*}

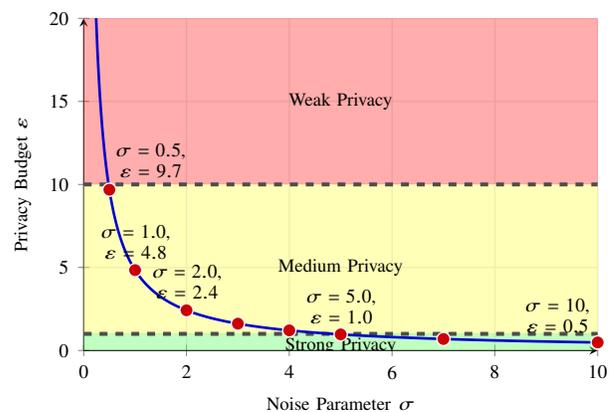
\begin{figure}[htbp]
\centering
\begin{tikzpicture}
\begin{axis}[
    width=0.95\linewidth,
    height=6cm,
    xlabel={Noise Parameter $\sigma$},
    ylabel={Privacy Budget $\varepsilon$},
    xmin=0, xmax=10,
    ymin=0, ymax=20,
    xtick={0,2,4,6,8,10},
    ytick={0,5,10,...,20},
    grid=major,
    grid style={line width=0.5pt, draw=gray!20},
    axis lines=left,
    scaled ticks=false,
    tick label style={font=\scriptsize},
    label style={font=\scriptsize},
    title style={font=\scriptsize},
]

\fill[red!80,opacity=0.4] (axis cs:0,10) rectangle (axis cs:10,50);
\fill[lemon!80,opacity=0.4] (axis cs:0,1) rectangle (axis cs:10,10);
\fill[green!60,opacity=0.4] (axis cs:0,0) rectangle (axis cs:10,1);

\draw[dashed, black!70, line width=1.5pt] (axis cs:0,1) -- (axis cs:10,1);
\draw[dashed, black!70, line width=1.5pt] (axis cs:0,10) -- (axis cs:10,10);

\addplot[
    domain=0.1:10,
    samples=300,
    smooth,
    line width=1pt,
    color=blue!80!black,
] {4.84/x};

\addplot[
    mark=*,
    mark size=2.5pt,
    mark options={fill=red!80!black, draw=white, line width=0.5pt},
    only marks
] coordinates {
    (0.5, 9.68)   
    (1.0, 4.84)   
    (2.0, 2.42)   
    (3.0, 1.613)  
    (4.0, 1.21)   
    (5.0, 0.968)  
    (7.0, 0.691)  
    (10.0, 0.484) 
};

\node[anchor=center, font=\scriptsize, text=black,  inner sep=4pt,  line width=1pt] at (axis cs:5, 15) {Weak Privacy};

\node[anchor=center, font=\scriptsize, text=black,  inner sep=4pt,  line width=1pt] at (axis cs:5, 5) {Medium Privacy};

\node[anchor=center, font=\scriptsize, text=black,  inner sep=4pt,  line width=1pt] at (axis cs:5, 0.3) {Strong Privacy};

\node[anchor=south west, font=\scriptsize, rounded corners=1pt, inner sep=1pt,text width=1cm, align=center] at (axis cs:0.5, 10.2) {$\sigma=0.5$, $\varepsilon=9.7$};

\node[anchor=south, font=\scriptsize, rounded corners=1pt, inner sep=1pt, text width=1cm, align=center] at (axis cs:1.0, 5.3) {$\sigma=1.0$, $\varepsilon=4.8$};

\node[anchor=south, font=\scriptsize, rounded corners=1pt, inner sep=1pt, text width=1cm, align=center] at (axis cs:2.0, 2.9) {$\sigma=2.0$, $\varepsilon=2.4$};

\node[anchor=south, font=\scriptsize, rounded corners=1pt, inner sep=1pt,text width=1cm, align=center] at (axis cs:5.0, 1.4) {$\sigma=5.0$, $\varepsilon=1.0$};

\node[anchor=south east, font=\scriptsize, rounded corners=1pt, inner sep=1pt,text width=1cm, align=center] at (axis cs:10.0, 0.9) {$\sigma=10$, $\varepsilon=0.5$};

\node[anchor=west, font=\scriptsize, text=black!80, fill=white, fill opacity=0.8, inner sep=2pt, rounded corners=1pt] at (axis cs:10.2, 1) {Strong/Medium boundary};

\node[anchor=west, font=\scriptsize, text=black!80, fill=white, fill opacity=0.8, inner sep=2pt, rounded corners=1pt] at (axis cs:10.2, 10) {Medium/Weak boundary};

\node[anchor=north west, font=\scriptsize, fill=white, fill opacity=0.9, draw=black!20, rounded corners=2pt, inner sep=4pt] at (axis cs:0.2, 48) {$\varepsilon = \frac{\Delta\sqrt{2\ln(1.25/\delta)}}{\sigma}$, $\Delta=1.0$, $\delta=10^{-5}$};

\end{axis}
\end{tikzpicture}
\caption{The inverse relationship between noise parameter $\sigma$ and privacy budget $\varepsilon$ in differential privacy, divided into three distinct zones: {Strong Privacy} ($\varepsilon \leq 1$), {Medium Privacy} ($1 < \varepsilon \leq 10$), and {Weak Privacy} ($\varepsilon > 10$). Higher noise values provide stronger privacy guarantees but reduce data utility.}
\label{fig:differential_privacy_tradeoff}
\end{figure}

Figure~\ref{fig:differential_privacy_tradeoff} illustrates the fundamental inverse relationship between noise parameters and privacy budgets, establishing three distinct privacy zones for practical deployment guidance.  This trade-off manifests differently across feature modalities due to differences in user population size, feature dimensionality, and semantic expressiveness. Contextual features exhibit the highest robustness to DP noise. With over 97,000 users and high-cardinality attributes such as IP geolocation, timezone, device characteristics, and browser metadata, individual contributions become statistically diluted. Consequently, contextual models tolerate strong noise levels (e.g., $\sigma \geq 5.0$, $\varepsilon \leq 1.0$) with minimal utility loss, making them well-suited for deployments under strict regulatory privacy budgets.

In contrast, keystroke dynamics are highly sensitive to noise due to their low-dimensional temporal structure and small user base (51 individuals), where each user’s data has a large influence on model behavior. As a result, even moderate noise levels (e.g., $\sigma = 0.3$) lead to noticeable performance degradation, and strong privacy guarantees become impractical. Mouse dynamics represent a middle ground: while the dataset includes only 10 users, the richer behavioral signals---such as movement trajectories, velocity, and click timing---enable moderate resilience to noise. At $\sigma = 1.0$ (yielding $\varepsilon \approx 5.0$), mouse-based models retain acceptable accuracy. Overall, contextual features are the most DP-resilient, followed by mouse dynamics, with keystroke features being the most vulnerable---highlighting the importance of modality-aware DP calibration in \acrshort{FLRBA2} for effective and privacy-preserving authentication.

\begin{figure*}[ht]
\centering
\scriptsize

\begin{subfigure}[t]{0.32\textwidth}
\centering
\resizebox{\linewidth}{!}{%
\begin{tikzpicture}
\begin{axis}[
    height=4.5cm,
    width=\linewidth,
    scale only axis,
    xlabel={$\sigma$ (Gaussian Noise)},
    ylabel={Performance Score},
    ymin=0.4, ymax=1.0,
    xmin=0, xmax=1,
    xtick={0,0.2,...,1},
    ytick={0.4,0.5,...,1.0},
    tick label style={font=\scriptsize},
    label style={font=\scriptsize},
    legend style={font=\scriptsize, at={(0.05,0.4)}, anchor=north west, legend columns=1},
    grid=both,
    major grid style={line width=.2pt,draw=gray!20},
    mark size=2pt
]
\addplot+[mark=o, color=blue] coordinates {
    (0.00, 0.7265) (0.01, 0.7286) (0.02, 0.7286) (0.05, 0.7307)
    (0.10, 0.7398) (0.15, 0.7419) (0.20, 0.7426) (0.30, 0.7645)
    (0.50, 0.7910) (0.70, 0.8067) (1.00, 0.8246)
};
\addlegendentry{Precision}
\addplot+[mark=square*, color=red] coordinates {
    (0.00, 0.7944) (0.01, 0.7944) (0.02, 0.7944) (0.05, 0.7944)
    (0.10, 0.7882) (0.15, 0.7882) (0.20, 0.7819) (0.30, 0.7788)
    (0.50, 0.7664) (0.70, 0.7539) (1.00, 0.7321)
};
\addlegendentry{Recall}
\addplot+[mark=triangle*, color=green!60!black] coordinates {
    (0.00, 0.7589) (0.01, 0.7601) (0.02, 0.7601) (0.05, 0.7612)
    (0.10, 0.7632) (0.15, 0.7644) (0.20, 0.7618) (0.30, 0.7716)
    (0.50, 0.7785) (0.70, 0.7794) (1.00, 0.7756)
};
\addlegendentry{F1-score}
\addplot+[mark=diamond*, color=black] coordinates {
    (0.00, 0.7944) (0.01, 0.7944) (0.02, 0.7944) (0.05, 0.7944)
    (0.10, 0.7882) (0.15, 0.7882) (0.20, 0.7819) (0.30, 0.7788)
    (0.50, 0.7664) (0.70, 0.7539) (1.00, 0.7321)
};
\addlegendentry{Accuracy}
\end{axis}
\end{tikzpicture}}
\caption{Keystroke}
\label{fig:dp_keystroke}
\end{subfigure}
\hfill
\begin{subfigure}[t]{0.32\textwidth}
\centering
\resizebox{\linewidth}{!}{%
\begin{tikzpicture}
\begin{axis}[
    height=4.5cm,
    width=\linewidth,
    scale only axis,
    xlabel={$\sigma$ (Gaussian Noise)},
    ylabel={Performance Score},
    ymin=0.4, ymax=1.0,
    xmin=0, xmax=2,
    xtick={0,0.2,...,2},
    ytick={0.4,0.5,...,1.0},
    tick label style={font=\scriptsize},
    label style={font=\scriptsize},
    legend style={font=\scriptsize, at={(0.05,0.4)}, anchor=north west, legend columns=1},
    grid=both,
    major grid style={line width=.2pt,draw=gray!20},
    mark size=2pt
]
\addplot+[mark=o, color=blue] coordinates {
    (0.00, 0.8947) (0.02, 0.8947)  (0.1, 0.8947) (0.2, 0.8947) (0.3, 0.8947) (1, 0.8947) (1.3, 0.8889)(1.5, 0.8889) (1.7, 0.8889) (1.8, 0.8889) (2, 0.8824)
};
\addlegendentry{Precision}
\addplot+[mark=square*, color=red] coordinates {
    (0.00, 0.7083) (0.02, 0.7083)  (0.1, 0.7083) (0.2, 0.7083)(0.3, 0.7083)(1, 0.7083) (1.3,0.6667) (1.5, 0.6667) (1.7, 0.6667)(1.8, 0.6667)  (2, 0.625)
};
\addlegendentry{Recall}
\addplot+[mark=triangle*, color=green!60!black] coordinates {
    (0.00, 0.7907) (0.02, 0.7907)  (0.1, 0.7907) (0.2, 0.7907)(0.3, 0.7907)(1, 0.7907) (1.3,0.7619)(1.5,0.7619) (1.7,0.7619)(1.8,0.7619) (2, 0.7317)
};
\addlegendentry{F1-score}
\addplot+[mark=diamond*, color=black] coordinates {
    (0.00, 0.7083) (0.02, 0.7083)  (0.1, 0.7083) (0.2, 0.7083)(0.3, 0.7083)(1, 0.7083) (1.3, 0.6667) (1.5, 0.6667) (1.7, 0.6667)(1.8, 0.6667) (2, 0.625)
};
\addlegendentry{Accuracy}
\end{axis}
\end{tikzpicture}}
\caption{Mouse}
\label{fig:dp_mouse}
\end{subfigure}
\hfill
\begin{subfigure}[t]{0.32\textwidth}
\centering
\resizebox{\linewidth}{!}{%
\begin{tikzpicture}

\begin{axis}[
    height=4.5cm,
    width=\linewidth,
    scale only axis,
    xlabel={$\sigma$ (Gaussian Noise)},
    ylabel={Performance Score},
    ymin=0.4, ymax=1.0,
    xmin=0, xmax=8,
    xtick={0,2,...,8},
    ytick={0.4,0.5,...,1.0},
    tick label style={font=\scriptsize},
    label style={font=\scriptsize},
    legend style={font=\scriptsize, at={(0.05,0.4)}, anchor=north west, legend columns=1},
    grid=both,
    major grid style={line width=.2pt,draw=gray!20},
    mark size=2pt
]

\addplot+[mark=o, color=blue] coordinates {
    (0.0, 0.7323) (0.1, 0.7323) (0.2, 0.7322) (0.5, 0.7328)
    (1.0, 0.7324) (1.5, 0.7338) (2.0, 0.7324) (3.0, 0.7328)
    (5.0, 0.7280) (8.0, 0.7326)
};
\addlegendentry{Precision}
\addplot+[mark=square*, color=red] coordinates {
    (0.0, 0.9609) (0.1, 0.9609) (0.2, 0.9609) (0.5, 0.9609)
    (1.0, 0.9608) (1.5, 0.9601) (2.0, 0.9607) (3.0, 0.9607)
    (5.0, 0.9632) (8.0, 0.9608)
};
\addlegendentry{Recall}
\addplot+[mark=triangle*, color=green!60!black] coordinates {
    (0.0, 0.8031) (0.1, 0.8031) (0.2, 0.8030) (0.5, 0.8033)
    (1.0, 0.8031) (1.5, 0.8037) (2.0, 0.8031) (3.0, 0.8033)
    (5.0, 0.8015) (8.0, 0.8032)
};
\addlegendentry{F1-score}
\addplot+[mark=diamond*, color=black] coordinates {
    (0.0, 0.9698) (0.1, 0.9698) (0.2, 0.9698) (0.5, 0.9698)
    (1.0, 0.9698) (1.5, 0.9692) (2.0, 0.9697) (3.0, 0.9697)
    (5.0, 0.9715) (8.0, 0.9698)
};
\addlegendentry{Accuracy}
\end{axis}
\end{tikzpicture}}
\caption{Contextual}
\label{fig:dp_contextual}
\end{subfigure}
\vspace{0.5em}
\caption{Privacy-utility trade-offs across authentication modalities. Contextual features maintain robust performance under strong DP constraints, while behavioral biometrics require careful noise calibration.}\vspace{-0.5em}
\label{fig:dp_comparison}
\end{figure*}
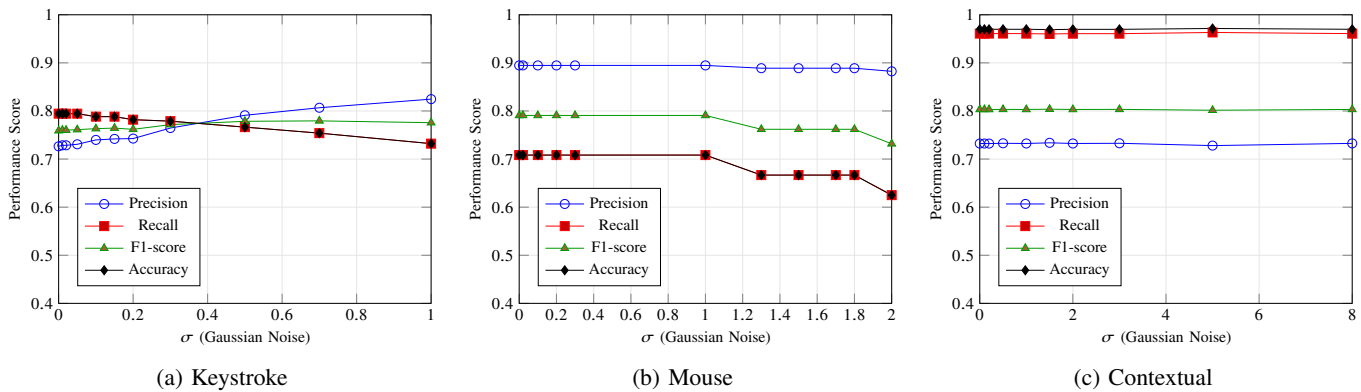

\subsection{Practical Implications}
\label{subsec:deployment}
Our evaluation yields evidence-driven deployment strategies for \acrshort{FLRBA2} tailored to varying privacy constraints, threat models, and operational needs. For high-privacy environments (e.g., GDPR-compliant systems, healthcare, public-sector deployments), we recommend exclusive contextual features under aggressive \gls{DP} ($\boldsymbol{\sigma} \geq 5.0$, yielding $\boldsymbol{\varepsilon} \leq 1.0$). These high-entropy attributes---user-agent strings, location tags, device metadata---maintain 0.9698 accuracy and 0.8031 F1-score under substantial noise, supported by high-cardinality features and over 97,000 users.
Enterprise settings requiring balanced privacy-security benefit from hybrid keystroke-contextual configurations under moderate noise ($\boldsymbol{\sigma} = 1.0$, $\boldsymbol{\varepsilon} \approx 5.0$), providing layered defense against behavioral mimicry and contextual spoofing while achieving 0.7943 and 0.9698 accuracy respectively. 
For scenarios prioritizing low false positives, such as fraud prevention or financial authentication, mouse dynamics offer the highest precision (0.8947) among all modalities. However, their moderate sensitivity to \gls{DP} noise and smaller user population necessitate conservative settings ($\boldsymbol{\sigma} \leq 1.0$). Practitioners should also account for their practical behavioral throughput when evaluating suitability for real-time enforcement.
\acrshort{FLRBA2} supports adaptive privacy tuning, enabling real-time noise parameter adjustment for changing regulatory or threat conditions, ensuring progressive strengthening of privacy guarantees while maintaining functional authentication. The framework provides practitioners flexible, privacy-preserving strategies aligned with real-world deployment constraints and performance goals.

\section{Conclusion}
\label{sec:conclusion}

We introduced \acrshort{FLRBA2}, a federated learning framework addressing dual challenges of privacy preservation and data heterogeneity in decentralized authentication. By transforming contextual, behavioral, biometric, and interaction-based features into standardized similarity representations, the framework enables interpretable, personalized risk modeling and effective federated learning over {Non-IID} data without centralizing raw inputs while mitigating cold-start issues.
\acrshort{FLRBA2} achieves strong high-risk detection performance across modalities: contextual features (F1-score 0.8031, recall 0.9609), mouse dynamics (F1-score 0.7907, precision 0.8947), and keystroke sequences (F1-score 0.7589, accuracy 0.7943). The system integrates \gls{DP} for noise-injected updates, message authentication codes for integrity and replay protection, and formally verified game-based security proofs under the random oracle model establishing guarantees for adaptive authentication, confidentiality, and update integrity.
Our modality-aware analysis reveals non-uniform privacy-utility trade-offs, reinforcing the importance of tailored deployment strategies. \acrshort{FLRBA2} offers a scalable, secure, privacy-preserving solution for continuous risk-based authentication, with promising research directions in dynamic user modeling, adversarial robustness, and trusted edge integration.

\bibliographystyle{IEEEtran}
\bibliography{references}

\appendices

\end{document}